\documentclass[reprint, superscriptaddress, twocolumn, amsmath, amssymb, aps, prb]{revtex4-2}
\usepackage{graphicx}
\usepackage{dcolumn}
\usepackage{bm}
\usepackage{placeins}
\usepackage{appendix}
\usepackage{upgreek}

\usepackage{verbatim}
\usepackage[usenames,dvipsnames]{xcolor}
 \usepackage{amsmath}
 \usepackage{amsfonts}
 \usepackage{amssymb}
 \usepackage[colorlinks=true,citecolor=blue,linkcolor=red]{hyperref}

 \usepackage{bbold}     
 \usepackage[makeroom]{cancel}  
 \usepackage{multirow}    
 \usepackage[normalem]{ulem}        
 \usepackage{array}
 \usepackage{pdfpages}
 \usepackage{lipsum}
\makeatletter
 \AtBeginDocument{\let\LS@rot\@undefined}
 \makeatother
\begin{document}

\title{Coulomb blockade in open superconducting islands on InAs nanowires}

\author{Huading Song}
\email{equal contribution}
\email{songhd@baqis.ac.cn}
\affiliation{Beijing Academy of Quantum Information Sciences, Beijing 100193, China}

\author{Zhaoyu Wang}
\email{equal contribution}
\affiliation{State Key Laboratory of Low Dimensional Quantum Physics, Department of Physics, Tsinghua University, Beijing 100084, China}

\author{Dong Pan}
\email{equal contribution}
\affiliation{State Key Laboratory of Superlattices and Microstructures, Institute of Semiconductors, Chinese Academy of Sciences, P.O. Box 912, Beijing 100083, China}

\author{Jiaye Xu}
\affiliation{State Key Laboratory of Low Dimensional Quantum Physics, Department of Physics, Tsinghua University, Beijing 100084, China}

\author{Yuqing Wang}
\affiliation{Beijing Academy of Quantum Information Sciences, Beijing 100193, China}

\author{Zhan Cao}
\affiliation{Beijing Academy of Quantum Information Sciences, Beijing 100193, China}

\author{Dong E. Liu}
\affiliation{State Key Laboratory of Low Dimensional Quantum Physics, Department of Physics, Tsinghua University, Beijing 100084, China}
\affiliation{Beijing Academy of Quantum Information Sciences, Beijing 100193, China}
\affiliation{Frontier Science Center for Quantum Information, Beijing 100084, China}
\affiliation{Hefei National Laboratory, Hefei 230088, China}

\author{Ke He}
\affiliation{State Key Laboratory of Low Dimensional Quantum Physics, Department of Physics, Tsinghua University, Beijing 100084, China}
\affiliation{Beijing Academy of Quantum Information Sciences, Beijing 100193, China}
\affiliation{Frontier Science Center for Quantum Information, Beijing 100084, China}
\affiliation{Hefei National Laboratory, Hefei 230088, China}

\author{Runan Shang}
\email{shangrn@baqis.ac.cn}
\affiliation{Beijing Academy of Quantum Information Sciences, Beijing 100193, China}
\affiliation{Hefei National Laboratory, Hefei 230088, China}

\author{Jianhua Zhao}
\affiliation{State Key Laboratory of Superlattices and Microstructures, Institute of Semiconductors, Chinese Academy of Sciences, P.O. Box 912, Beijing 100083, China}

\author{Hao Zhang}
\email{hzquantum@mail.tsinghua.edu.cn}
\affiliation{State Key Laboratory of Low Dimensional Quantum Physics, Department of Physics, Tsinghua University, Beijing 100084, China}
\affiliation{Beijing Academy of Quantum Information Sciences, Beijing 100193, China}
\affiliation{Frontier Science Center for Quantum Information, Beijing 100084, China}


\begin{abstract}

Electrons in closed systems can exhibit Coulomb blockade (CB) oscillations due to charge quantization. Here, we report CB oscillations in aluminum superconducting islands on InAs nanowires in the open regime. The Al island is connected to the source/drain leads through two contacts: One is fully transmitting while the other is tuned into the tunneling regime. This device configuration is typical for tunneling spectroscopy where charging energy is generally considered negligible. The oscillation periods are 2$e$ or 1$e$, depending on the gate settings. A magnetic field can induce the 2$e$ to 1$e$ transition. Our result is reminiscent of the ``mesoscopic Coulomb blockade'' in open quantum dots caused by electron interference. 

\end{abstract}

\maketitle

Quantum dots are closed mesoscopic systems where electrons are weakly coupled to the outside reservoirs via tunneling contacts \cite{RMP_QD}. In the weak tunneling regime, the transmissions through the contacts are small ($\ll$ 1), resulting in well-localized electrons. Coulomb interactions dominate transport, manifesting as Coulomb blockade (CB) oscillations in device conductance. In the open regime where contacts are fully transmitting, charge quantization and CB should disappear. An interesting intermediate case arises when one point contact is fully transmitting while the other remains in the tunneling regime. In this scenario, CB oscillations can persist due to electron interference, which reduces the transmission of the open contact and traps electrons on the dot \cite{Marcus_open_QD_1998}. This effect, known as mesoscopic CB, can also be observed in open quantum dots with two fully transmitting leads \cite{DGG_open_QD_2011}. The mesoscopic CB oscillations can be suppressed by a magnetic field which destroys the interference between time-reversal paths. However, the CB has not been reported in superconducting island with fully transmitting contact, a superconducting version of open quantum dot.

In this work, we report a phenomena similar to the mesoscopic CB in aluminum superconducting islands coupled to InAs nanowires. These hybrid semiconductor-superconductor nanowires have been intensively studied as a promising material platform for realizing Majorana zero modes \cite{Lutchyn2010, Oreg2010, NextSteps, Prada2020, Leo_Perspective}. Two typical device configurations are tunneling spectroscopy \cite{Mourik, Deng2016, Gul2018, Song2022, WangZhaoyu, MS_2023, Delft_Kitaev} and island CB transport \cite{Albrecht}. In the former configuration, the superconductor is grounded, i.e., directly contacting the lead reservoir, resulting in negligible charging energy. In the latter case (CB regime), the superconducting island is tunnel-coupled to the reservoirs through the semiconductor nanowire, localizing quasiparticles and Cooper pairs on the island. Although the device configuration in this work corresponds to the former case, we observe CB oscillations with periods of $2e$ and $1e$. We attribute this phenomenon to a mechanism similar to mesoscopic CB, where coherent Andreev reflections at the open contact facilitate charge localization on the island. These CB oscillations can be diminished by a magnetic field, which suppresses the Andreev processes.      

 \begin{figure}[!ht]
 \includegraphics[width=\columnwidth]{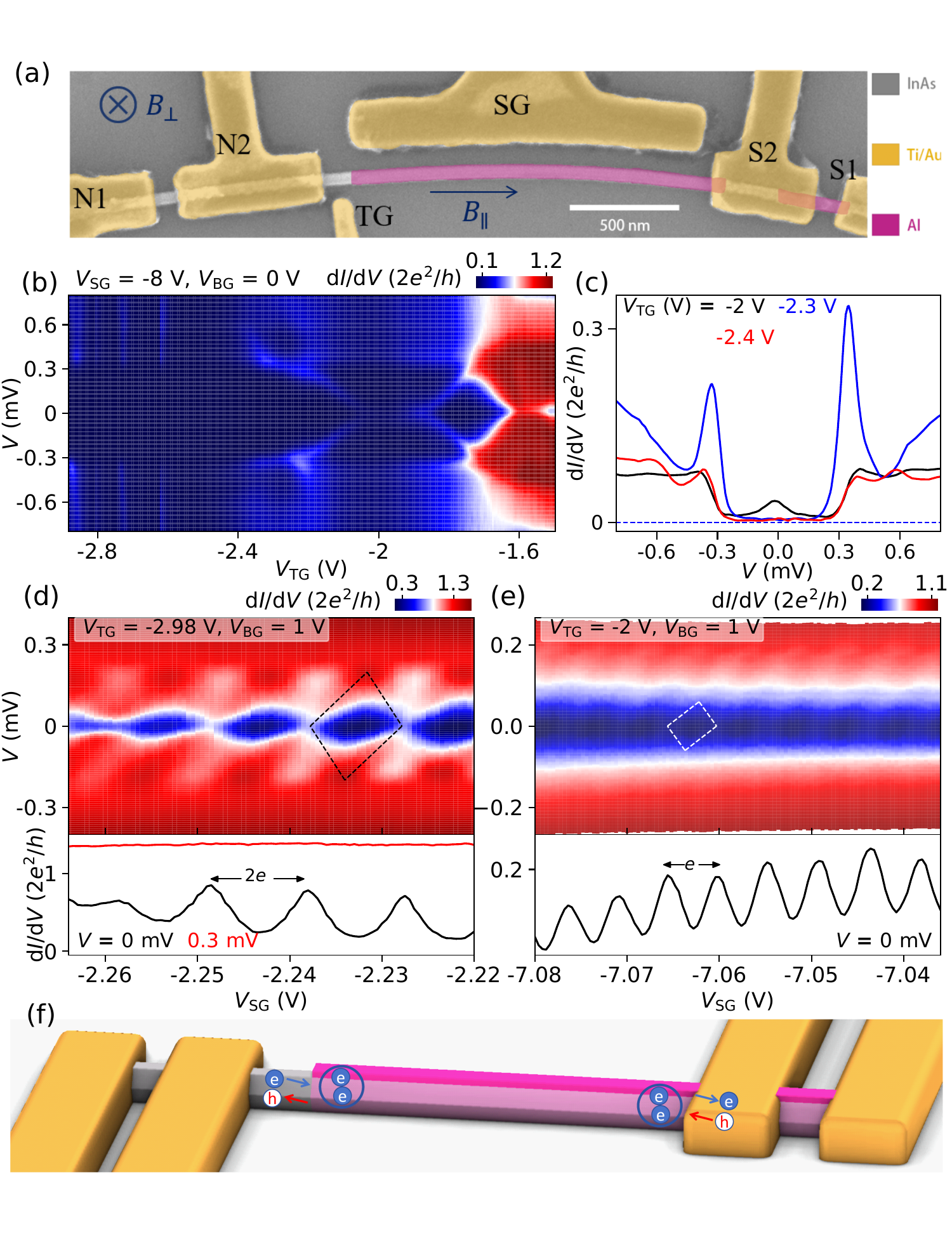}
 \caption{\label{Fig1}(a) False-colored scanning electron microscope (SEM) image of device A. The Al shell is in pink. The contacts and gates are Ti/Au with a thickness of 5/70 nm (yellow). The substrate is p-doped Si covered by 300-nm-thick SiO$_2$. Orientations of parallel and perpendicular magnetic fields ($B_{\parallel}$ and $B_{\perp}$) are sketched. (b) Hard gap tunneling spectroscopy at $B$ = 0 T. $V_\mathrm{TG}$ = -8 V. (c) Vertical line cuts from (a) at $V_\mathrm{TG}$ = -2 V, -2.3 V, and -2.4 V, respectively. (d,e) d$I$/d$V$ as a function of $V_\mathrm{SG}$ and $V$ at $B$ = 0 T, demonstrating $2e$ (d) and $1e$ (e) Coulomb blockade oscillations, respectively. The black and white dashed lines mark the diamond shapes of the $2e$ and $1e$ blockades. The lower panels show line cuts at $V$ = 0 mV (black) and 0.3 mV (red). (f) Device schematic with the process of Andreev reflection sketched.}
 \end{figure}

Figure 1(a) shows a scanning electron micrograph (SEM) of the device (device A). The InAs-Al nanowire was grown in situ using molecular beam epitaxy. For detailed information of growth, we refer to Ref. \cite{PanCPL}. The diameter of InAs is relatively thin ($\sim$ 46 nm). The Al shell is 7 nm thick. The four electrical contacts are labeled as N1, N2, S1, and S2. The device has two side gates, TG and SG, and one global back gate, BG. The substrate is p-doped Si (BG) covered by 300-nm-thick SiO$_2$. The measurement was carried out in a dilution fridge with a base temperature of $\sim$ 20 mK using four-terminal measurement \cite{WangZhaoyu}. A bias voltage, along with a lock-in excitation, was applied to contact N1, while the corresponding current $I$ and d$I$ were drained from S1. A voltage meter measured $V$ and d$V$ between N2 and S2, allowing us to exclude contact resistance through this four-terminal configuration.

In Fig. 1(b), the device was tuned into the tunneling regime. Differential conductance, d$I$/d$V$, was measured as a function of $V$ and the tunnel gate voltage $V_{\text{TG}}$. The magnetic field ($B$) was set to zero. A hard gap is observed, as indicated by the blue line cut in Fig. 1(c), with a gap size of $\sim$ 0.36 meV. Gate-tunable subgap states, i.e. Andreev bound states \cite{Silvano2014}, can also be revealed, see the black line cut in Fig. 1(c). The hard gap can sustain a parallel magnetic field ($B_{||}$) up to 3 T, see Fig. S1 in the Supplemental Material. The orientations of $B_{||}$ and $B_{\perp}$ are illustrated in Fig. 1(a). 

Figures 1(d) and 1(e) depict the main result of the paper: At specific gate settings (labeled in the figure panels), the zero-bias conductance (lower panels) exhibits periodic oscillations while the corresponding bias spectroscopy (upper panels) reveals diamond-like shapes (dashed lines). The oscillation periods in $V_{\text{SG}}$ are 10 mV for Fig. 1(d) and 5 mV for Fig. 1(e), respectively. This difference by a factor of two suggests the $2e$ and $1e$ periodic nature of the two oscillations, as indicated in the lower panels. The diamond size in bias $V$ is $\sim$ 0.2 mV for Fig. 1(d) and $\sim$ 0.057 mV for Fig. 1(e). Assuming the charging energy $E_c = e^2/2C$, the $2e$ and $1e$ diamond sizes correspond to $8E_c$ and $2E_c$, respectively. This ratio is also roughly consistent with the diamond sizes in Figs. 1(d-e), from which we estimate $E_c$ $\sim$ 25-29 $\upmu$eV.  For additional bias spectroscopy scans and a summary of oscillation periods, see Fig. S1 and Table S1 in Supplemental Material. 

This charging energy is orders of magnitude smaller than that of a typical quantum dot formed in the tunnel barrier region of the nanowire, yet it is consistent with the CB observed in a InAs-Al island \cite{PanCPL}. We have also confirmed in Fig. S2 that the oscillations are still present if using N2 and S2 as the leads in a two-terminal measurement (N1 and S1 were kept floated). Thus, the charging energy likely arises from the InAs-Al island between N2 and S2. This is slightly counter-intuitive since the island is in principle grounded by the S2 contact. We note that these oscillations were also present in our previous device \cite{WangZhaoyu} but not investigated systematically. The charge localization is less likely resulting from the poor contacting of S2, due to the low contact resistance ($\sim$ 250 $\Omega$) extracted from the comparison between four-terminal and two-terminal measurements. A possible mechanism for the charge localization is the mesoscopic CB, previously observed in open quantum dots \cite{Marcus_open_QD_1998, DGG_open_QD_2011, Glazman_CB}. In this scenario, the left contact of the island operates in the tunneling regime while the right contact is in the open regime. The key difference here lies in the contact couplings, which are mediated by Andreev reflections in both contacts rather than normal scattering. It is important to note that the superconducting island is phase coherent. The quantum interference between Andreev reflections localizes the charges on the island, an Andreev version of mesoscopic CB. The oscillations are generally observed in thin nanowire devices (based on our results, $diameter < 60 nm$), highlighting the crucial role of (minimal) subband occupation. However, the detailed mechanism underlying the gate-dependent transition from 2$e$ to 1$e$ oscillation requires further investigation. Figure S3 shows an intermediate region between the 2$e$ and 1$e$ oscillation regions, where the spacing between the oscillation peaks is non-uniform. Emergence of sub-gap states or quasi-particle poisoning in the InAs-Al island may cause this transition \cite{Albrecht_PRL, Albrecht}.

 \begin{figure*}[!ht]
 \includegraphics[width=\textwidth]{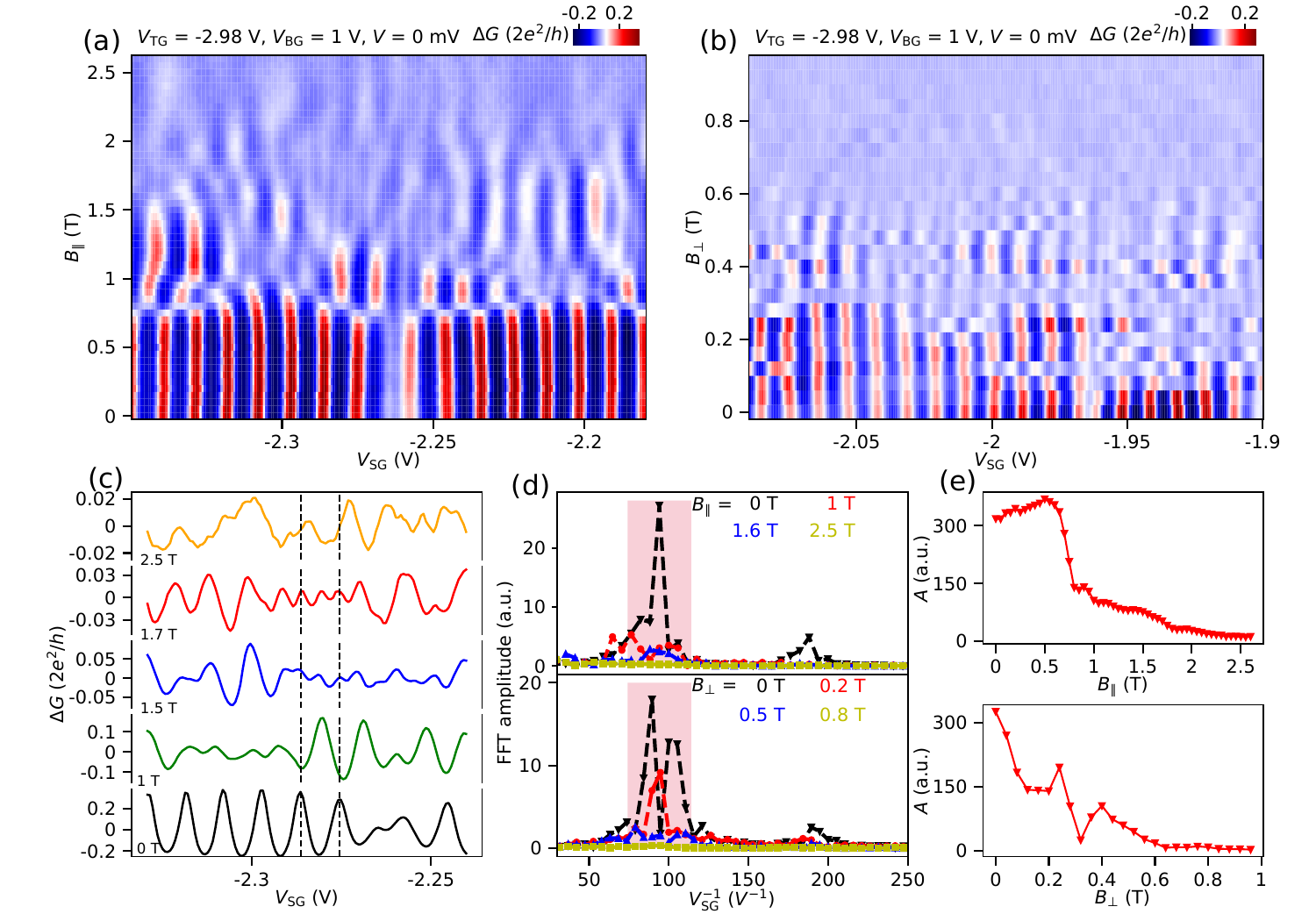}
 \caption{\label{Fig2}(a, b) $B_{\parallel}$ and $B_{\perp}$ dependence of the oscillations, respectively.  $\Delta G$ is the variation of d$I$/d$V$. (c) Line cuts from (a) at $B_{\parallel}$ = 0, 1, 1.5, 1.7, and 2.3 T, respectively. The dashed lines indicate the oscillation peaks at 0 T and dips at 1 T. At higher fields, the oscillation period between the two dashed lines is reduced by half. (d) FFT analysis of (a) and (b), shown in the upper and lower panels, respectively. The 2$e$-periodic oscillations correspond to a FFT peak at $\sim$ 94 $ V^{-1}$. (e) Oscillation amplitudes ($A$) as function of $B$, extracted from the FFT of (a) and (b), respectively.}
\end{figure*}

We then study the magnetic field dependence of these oscillations. Figure 2(a) shows the $2e$ oscillations in a parallel field, $B_{||}$. For clarity, the background conductance was averaged and subtracted, ensuring that $\Delta G$ primarily reflects the oscillation signal. Figure S4 in the Supplemental Material shows the process of background subtraction. The $2e$ CB persists to $\sim$ 0.7 T, after which a $\pi$ phase shift occurs, see Fig. 2(c) for the line cuts. The dashed lines indicate the oscillation peaks at 0 T, which become dips at 1.0 T (the green curve), indicating this $\pi$ shift. At higher field ($\sim$ 1.5 T and 1.7 T), the oscillation period between the two dashed lines are halved (the blue and red curves), suggesting the $1e$ nature of the oscillations. In Fig. S5, we track the oscillation period as a function of $B_{||}$ for $V_{\text{SG}}$ between -2.295 V and -2.255 V. The $\pi$-phase shift and the $2e$ to $1e$ transition may be associated with a subgap state crossing zero energy as sweeping $B_{||}$. For $B_{||}$ higher than 2 T, the oscillations gradually disappears, due to the suppression of the superconductivity and Andreev reflection. The $B$ evolution in Fig. 2(a) further confirm the $2e$ nature of the oscillations in Fig. 1(d). 

Figure 2(b) illustrates the $B$ dependence of the $2e$ CB with the field direction perpendicular to the device substrate ($B_{\perp}$). For $B$ dependence of the $1e$ oscillations, we refer to Fig. S6 in Supplemental Material. The oscillations are suppressed at $B_{\perp}$ $\sim$ 0.6 T, consistent with the field value where the superconductivity is suppressed \cite{ZhangShan, WangZhichuan}. This further indicates that the CB oscillations originate from Andreev reflections.

To quantify the field dependence of the oscillations, we calculated the oscillation amplitude $A$ by performing fast Fourier transform (FFT) on $\Delta$G. Figure 2(d) shows the FFT spectrum at several fields of Fig. 2(a) and Fig. 2(b), respectively. The dominant peak near 94 V$^{-1}$ correspond to the $2e$ CB oscillations while the secondary peak near 190 V$^{-1}$ corresponds to the $1e$ component. $A$ was estimated by integrating this spectrum near the oscillation frequency over a integration window (the shaded area in Fig. 2(d)). For details of the integration, see Fig. S4. Figure 2(e) shows the evolution of the $2e$-oscillation amplitude ($A$) with magnetic field for Fig. 2(a) and 2(b), respectively. For $B_{||}$, the $2e$ amplitude remains relatively unchanged for fields below 0.5 T but is fully suppressed at 2.5 T. For $B_{\perp}$, the $2e$ CB is fully suppressed at 0.6 T. Figure S6 shows a similar analysis on $1e$ CB. For additional data on CB oscillations in devices A and B, see Figs. S7.

\begin{figure}[!ht]
 \includegraphics[width=\columnwidth]{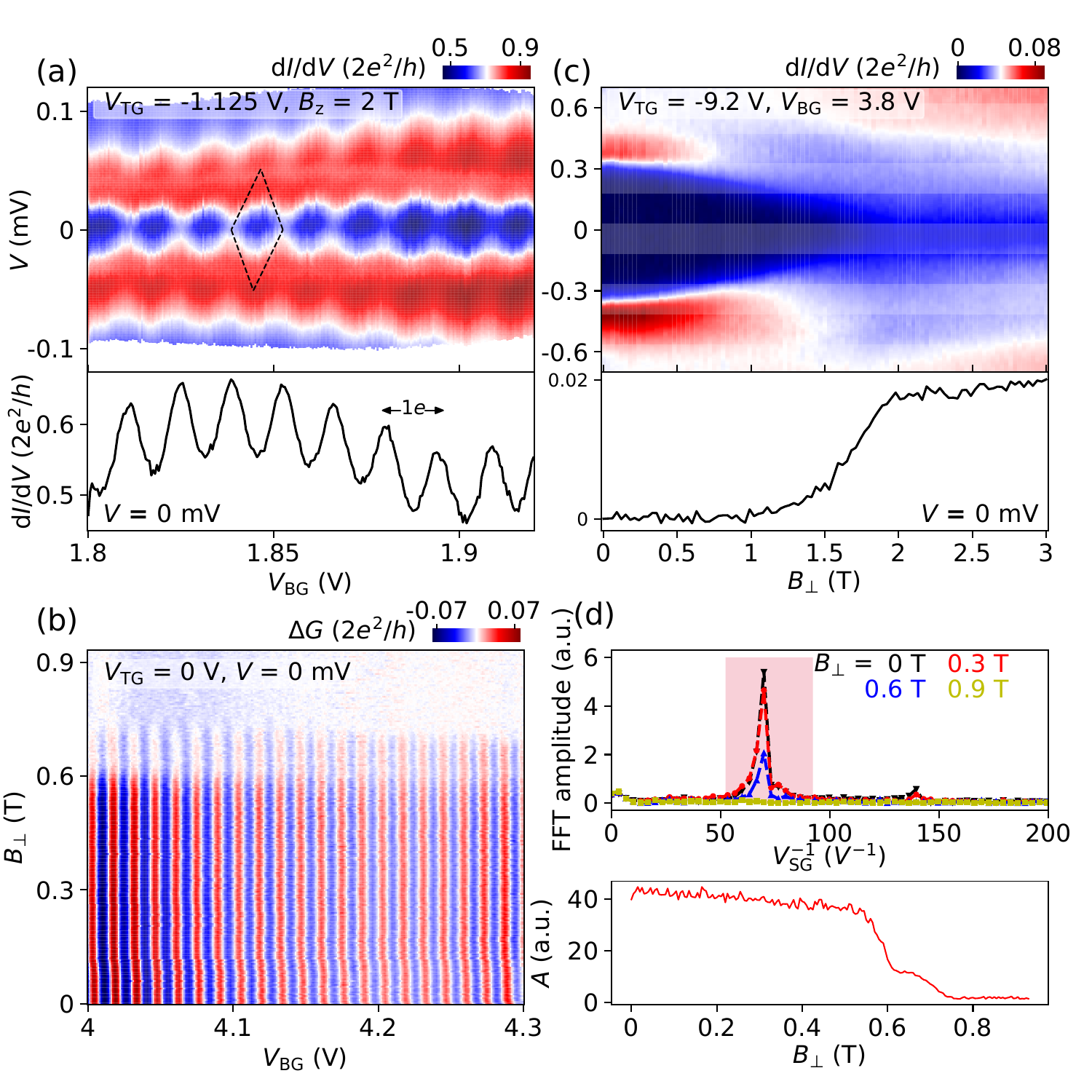}
\caption{\label{Fig3}(a) CB oscillations in a second device (device B). Upper panel, d$I$/d$V$ as a function of $V_{\text{SG}}$ and $V$ at $B_Z$  = 2 T. $B_Z$ is the $z$ axis of the vector magnet, which is 8 degrees off from the nanowire axis. Lower panel, zero-bias line cut, showing the $1e$ oscillations.  (b) $B_{\perp}$ dependence of the oscillations. (c) $B_{\perp}$ dependence of the superconducting gap of device B.  (d) FFT analysis of (b). }
\end{figure}

Next, we report CB oscillations observed in a second device (device B). This device has been extensively investigated in Ref. \cite{WangZhaoyu} in the context of quantized zero-bias peaks. The InAs has a diameter of 26 nm with the length of the Al region $\sim$ 1.1 $\upmu$m. In Ref. \cite{WangZhaoyu}, CB oscillations were observable and superimposed on the zero-bias peaks as a background. Here, we show their existence in the parameter regions without zero-bias peaks, see Fig. 3(a). The diamond shape is indicated by the dashed lines in the upper panel. The diamond size in bias is $\sim$ 50 $\upmu$V, corresponding to a $E_c$ of 25 $\upmu$eV if assuming the oscillations to be $1e$ periodic. This value is also consistent with the $E_c$ extracted from device A. Throughout the parameter space, we did not find $2e$ oscillations, possibly due to the quasi-particle poisoning introduced by the S2 contact.  

Figure 3(b) shows the $B_{\perp}$ dependence of the oscillations (background subtracted). Although in a different gate setting from Fig. 3(a), the oscillation periods in $V_{\text{BG}}$ are nearly identical, i.e. $\sim$ 13.8 mV. The oscillations are suppressed for $B_{\perp}$ above 0.6 T, lower than the critical field of the gap. In Fig. 3(c), we shows the $B_{\perp}$ dependence of the gap whose critical field is larger than 1.5 T. The zero-bias conductance (lower panel) exhibits an increasing trend for $B_{\perp}$ around 1 T, thus the softening of the gap at high field may suppress the Andreev process as well as the oscillations. Figure 3(d) shows the FFT analysis of Fig. 3(b). The $1e$ oscillations are revealed as the dominant peak near  $68 V^{-1}$. From the integration around this peak (the dahsed area), we extract the oscillation amplitude, $A$, which decreases sharply near 0.6 T, consistent with Fig. 3(b). 

\begin{figure}[!ht]
	\includegraphics[width=\columnwidth]{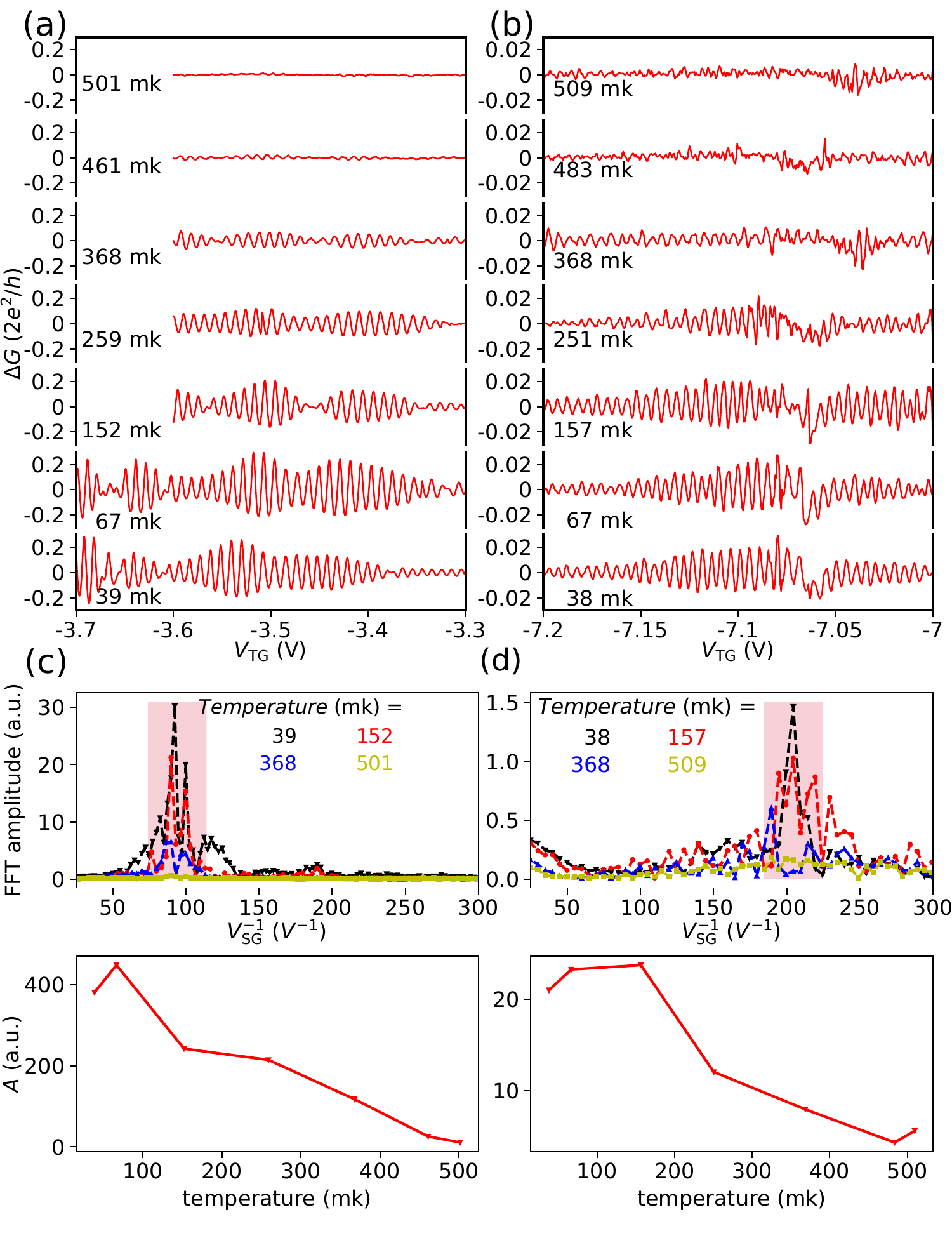}
	\caption{\label{Fig4}(a, b) Temperature dependence of the 2$e$ (a) and 1$e$ (b) oscillations in device A. $V_{\text{BG}} = 1 V$ $V_{\text{TG}} = -2 V$ for all curves. (c,d) Corresponding FFT analysis.}
\end{figure}
We then investigate the temperature dependence of the oscillations in device A. Figures 4(a,b) show the oscillation signals of 2$e$ and 1$e$ periods, respectively, measured at various temperatures (background subtracted). The oscillation amplitudes are roughly halved around 250 mK, and nearly zero at 500 mK. The upper panels in Figs. 4(c,d) show the FFT spectrum at several temperatures, while the lower panels illustrates the trend of amplitudes decay with increasing temperature.

To conclude, we have observed Coulomb blockade oscillations in InAs-Al nanowire devices in the open regime. The oscillation periods are $2e$ or $1e$, depending on specific gate settings. A magnetic field can suppress the oscillations as well as the superconductivity. These observations are reminiscent to the mesoscopic Coulomb blockade in an open quantum dot. The charge localization is likely caused by coherent Andreev reflections, despite of one contact being fully transmitting. Future study can explore the interplay between these oscillations and zero-energy subgap states, which may shed light on Andreev and Majorana related physics. Our findings also provide the possibility of achieving synchronized combined measurements (tunneling spectroscopy and island CB transport) in nanowire devices \cite{Valentini_Nature}.

Raw data for this Letter are openly available from the Zenodo repository \cite{oscillation_Zenodo}.

\textbf{Acknowledgment.} This work is supported by National Natural Science Foundation of China (Grants No. 12104053, No. 12374459, No. 61974138, No. 92065206, No. 12374158, No. 12074039, No. 11974198), the Innovation Program for Quantum Science and Technology (Grant 2021ZD0302400). D. P. also acknowledges the support from Youth Innovation Promotion Association, Chinese Academy of Sciences (No. 2017156, No. Y2021043).

\bibliography{small_oscillation}

\begin{thebibliography}{24}%
\makeatletter
\providecommand \@ifxundefined [1]{%
 \@ifx{#1\undefined}
}%
\providecommand \@ifnum [1]{%
 \ifnum #1\expandafter \@firstoftwo
 \else \expandafter \@secondoftwo
 \fi
}%
\providecommand \@ifx [1]{%
 \ifx #1\expandafter \@firstoftwo
 \else \expandafter \@secondoftwo
 \fi
}%
\providecommand \natexlab [1]{#1}%
\providecommand \enquote  [1]{``#1''}%
\providecommand \bibnamefont  [1]{#1}%
\providecommand \bibfnamefont [1]{#1}%
\providecommand \citenamefont [1]{#1}%
\providecommand \href@noop [0]{\@secondoftwo}%
\providecommand \href [0]{\begingroup \@sanitize@url \@href}%
\providecommand \@href[1]{\@@startlink{#1}\@@href}%
\providecommand \@@href[1]{\endgroup#1\@@endlink}%
\providecommand \@sanitize@url [0]{\catcode `\\12\catcode `\$12\catcode
  `\&12\catcode `\#12\catcode `\^12\catcode `\_12\catcode `\%12\relax}%
\providecommand \@@startlink[1]{}%
\providecommand \@@endlink[0]{}%
\providecommand \url  [0]{\begingroup\@sanitize@url \@url }%
\providecommand \@url [1]{\endgroup\@href {#1}{\urlprefix }}%
\providecommand \urlprefix  [0]{URL }%
\providecommand \Eprint [0]{\href }%
\providecommand \doibase [0]{https://doi.org/}%
\providecommand \selectlanguage [0]{\@gobble}%
\providecommand \bibinfo  [0]{\@secondoftwo}%
\providecommand \bibfield  [0]{\@secondoftwo}%
\providecommand \translation [1]{[#1]}%
\providecommand \BibitemOpen [0]{}%
\providecommand \bibitemStop [0]{}%
\providecommand \bibitemNoStop [0]{.\EOS\space}%
\providecommand \EOS [0]{\spacefactor3000\relax}%
\providecommand \BibitemShut  [1]{\csname bibitem#1\endcsname}%
\let\auto@bib@innerbib\@empty
\bibitem [{\citenamefont {Hanson}\ \emph {et~al.}(2007)\citenamefont {Hanson},
  \citenamefont {Kouwenhoven}, \citenamefont {Petta}, \citenamefont {Tarucha},\
  and\ \citenamefont {Vandersypen}}]{RMP_QD}%
  \BibitemOpen
  \bibfield  {author} {\bibinfo {author} {\bibfnamefont {R.}~\bibnamefont
  {Hanson}}, \bibinfo {author} {\bibfnamefont {L.~P.}\ \bibnamefont
  {Kouwenhoven}}, \bibinfo {author} {\bibfnamefont {J.~R.}\ \bibnamefont
  {Petta}}, \bibinfo {author} {\bibfnamefont {S.}~\bibnamefont {Tarucha}},\
  and\ \bibinfo {author} {\bibfnamefont {L.~M.~K.}\ \bibnamefont
  {Vandersypen}},\ }\bibfield  {title} {\bibinfo {title} {Spins in few-electron
  quantum dots},\ }\href {https://doi.org/10.1103/RevModPhys.79.1217}
  {\bibfield  {journal} {\bibinfo  {journal} {Rev. Mod. Phys.}\ }\textbf
  {\bibinfo {volume} {79}},\ \bibinfo {pages} {1217} (\bibinfo {year}
  {2007})}\BibitemShut {NoStop}%
\bibitem [{\citenamefont {Cronenwett}\ \emph {et~al.}(1998)\citenamefont
  {Cronenwett}, \citenamefont {Maurer}, \citenamefont {Patel}, \citenamefont
  {Marcus}, \citenamefont {Duru\"oz},\ and\ \citenamefont
  {Harris}}]{Marcus_open_QD_1998}%
  \BibitemOpen
  \bibfield  {author} {\bibinfo {author} {\bibfnamefont {S.~M.}\ \bibnamefont
  {Cronenwett}}, \bibinfo {author} {\bibfnamefont {S.~M.}\ \bibnamefont
  {Maurer}}, \bibinfo {author} {\bibfnamefont {S.~R.}\ \bibnamefont {Patel}},
  \bibinfo {author} {\bibfnamefont {C.~M.}\ \bibnamefont {Marcus}}, \bibinfo
  {author} {\bibfnamefont {C.~I.}\ \bibnamefont {Duru\"oz}},\ and\ \bibinfo
  {author} {\bibfnamefont {J.~S.}\ \bibnamefont {Harris}},\ }\bibfield  {title}
  {\bibinfo {title} {Mesoscopic {C}oulomb blockade in one-channel quantum
  dots},\ }\href {https://doi.org/10.1103/PhysRevLett.81.5904} {\bibfield
  {journal} {\bibinfo  {journal} {Phys. Rev. Lett.}\ }\textbf {\bibinfo
  {volume} {81}},\ \bibinfo {pages} {5904} (\bibinfo {year}
  {1998})}\BibitemShut {NoStop}%
\bibitem [{\citenamefont {Amasha}\ \emph {et~al.}(2011)\citenamefont {Amasha},
  \citenamefont {Rau}, \citenamefont {Grobis}, \citenamefont {Potok},
  \citenamefont {Shtrikman},\ and\ \citenamefont
  {Goldhaber-Gordon}}]{DGG_open_QD_2011}%
  \BibitemOpen
  \bibfield  {author} {\bibinfo {author} {\bibfnamefont {S.}~\bibnamefont
  {Amasha}}, \bibinfo {author} {\bibfnamefont {I.~G.}\ \bibnamefont {Rau}},
  \bibinfo {author} {\bibfnamefont {M.}~\bibnamefont {Grobis}}, \bibinfo
  {author} {\bibfnamefont {R.~M.}\ \bibnamefont {Potok}}, \bibinfo {author}
  {\bibfnamefont {H.}~\bibnamefont {Shtrikman}},\ and\ \bibinfo {author}
  {\bibfnamefont {D.}~\bibnamefont {Goldhaber-Gordon}},\ }\bibfield  {title}
  {\bibinfo {title} {Coulomb blockade in an open quantum dot},\ }\href
  {https://doi.org/10.1103/PhysRevLett.107.216804} {\bibfield  {journal}
  {\bibinfo  {journal} {Phys. Rev. Lett.}\ }\textbf {\bibinfo {volume} {107}},\
  \bibinfo {pages} {216804} (\bibinfo {year} {2011})}\BibitemShut {NoStop}%
\bibitem [{\citenamefont {Lutchyn}\ \emph {et~al.}(2010)\citenamefont
  {Lutchyn}, \citenamefont {Sau},\ and\ \citenamefont
  {Das~Sarma}}]{Lutchyn2010}%
  \BibitemOpen
  \bibfield  {author} {\bibinfo {author} {\bibfnamefont {R.~M.}\ \bibnamefont
  {Lutchyn}}, \bibinfo {author} {\bibfnamefont {J.~D.}\ \bibnamefont {Sau}},\
  and\ \bibinfo {author} {\bibfnamefont {S.}~\bibnamefont {Das~Sarma}},\
  }\bibfield  {title} {\bibinfo {title} {Majorana fermions and a topological
  phase transition in semiconductor-superconductor heterostructures},\ }\href
  {https://doi.org/10.1103/PhysRevLett.105.077001} {\bibfield  {journal}
  {\bibinfo  {journal} {Phys. Rev. Lett.}\ }\textbf {\bibinfo {volume} {105}},\
  \bibinfo {pages} {077001} (\bibinfo {year} {2010})}\BibitemShut {NoStop}%
\bibitem [{\citenamefont {Oreg}\ \emph {et~al.}(2010)\citenamefont {Oreg},
  \citenamefont {Refael},\ and\ \citenamefont {von Oppen}}]{Oreg2010}%
  \BibitemOpen
  \bibfield  {author} {\bibinfo {author} {\bibfnamefont {Y.}~\bibnamefont
  {Oreg}}, \bibinfo {author} {\bibfnamefont {G.}~\bibnamefont {Refael}},\ and\
  \bibinfo {author} {\bibfnamefont {F.}~\bibnamefont {von Oppen}},\ }\bibfield
  {title} {\bibinfo {title} {Helical liquids and {M}ajorana bound states in
  quantum wires},\ }\href {https://doi.org/10.1103/PhysRevLett.105.177002}
  {\bibfield  {journal} {\bibinfo  {journal} {Phys. Rev. Lett.}\ }\textbf
  {\bibinfo {volume} {105}},\ \bibinfo {pages} {177002} (\bibinfo {year}
  {2010})}\BibitemShut {NoStop}%
\bibitem [{\citenamefont {Zhang}\ \emph {et~al.}(2019)\citenamefont {Zhang},
  \citenamefont {Liu}, \citenamefont {Wimmer},\ and\ \citenamefont
  {Kouwenhoven}}]{NextSteps}%
  \BibitemOpen
  \bibfield  {author} {\bibinfo {author} {\bibfnamefont {H.}~\bibnamefont
  {Zhang}}, \bibinfo {author} {\bibfnamefont {D.~E.}\ \bibnamefont {Liu}},
  \bibinfo {author} {\bibfnamefont {M.}~\bibnamefont {Wimmer}},\ and\ \bibinfo
  {author} {\bibfnamefont {L.~P.}\ \bibnamefont {Kouwenhoven}},\ }\bibfield
  {title} {\bibinfo {title} {Next steps of quantum transport in {Majorana}
  nanowire devices},\ }\href {https://doi.org/10.1038/s41467-019-13133-1}
  {\bibfield  {journal} {\bibinfo  {journal} {Nature Communications}\ }\textbf
  {\bibinfo {volume} {10}},\ \bibinfo {pages} {5128} (\bibinfo {year}
  {2019})}\BibitemShut {NoStop}%
\bibitem [{\citenamefont {Prada}\ \emph {et~al.}(2020)\citenamefont {Prada},
  \citenamefont {San-Jose}, \citenamefont {de~Moor}, \citenamefont {Geresdi},
  \citenamefont {Lee}, \citenamefont {Klinovaja}, \citenamefont {Loss},
  \citenamefont {Nyg{\aa}rd}, \citenamefont {Aguado},\ and\ \citenamefont
  {Kouwenhoven}}]{Prada2020}%
  \BibitemOpen
  \bibfield  {author} {\bibinfo {author} {\bibfnamefont {E.}~\bibnamefont
  {Prada}}, \bibinfo {author} {\bibfnamefont {P.}~\bibnamefont {San-Jose}},
  \bibinfo {author} {\bibfnamefont {M.~W.}\ \bibnamefont {de~Moor}}, \bibinfo
  {author} {\bibfnamefont {A.}~\bibnamefont {Geresdi}}, \bibinfo {author}
  {\bibfnamefont {E.~J.}\ \bibnamefont {Lee}}, \bibinfo {author} {\bibfnamefont
  {J.}~\bibnamefont {Klinovaja}}, \bibinfo {author} {\bibfnamefont
  {D.}~\bibnamefont {Loss}}, \bibinfo {author} {\bibfnamefont {J.}~\bibnamefont
  {Nyg{\aa}rd}}, \bibinfo {author} {\bibfnamefont {R.}~\bibnamefont {Aguado}},\
  and\ \bibinfo {author} {\bibfnamefont {L.~P.}\ \bibnamefont {Kouwenhoven}},\
  }\bibfield  {title} {\bibinfo {title} {From {A}ndreev to {M}ajorana bound
  states in hybrid superconductor--semiconductor nanowires},\ }\href
  {https://doi.org/10.1038/s42254-020-0228-y} {\bibfield  {journal} {\bibinfo
  {journal} {Nature Reviews Physics}\ }\textbf {\bibinfo {volume} {2}},\
  \bibinfo {pages} {575} (\bibinfo {year} {2020})}\BibitemShut {NoStop}%
\bibitem [{\citenamefont {Kouwenhoven}(2024)}]{Leo_Perspective}%
  \BibitemOpen
  \bibfield  {author} {\bibinfo {author} {\bibfnamefont {L.~P.}\ \bibnamefont
  {Kouwenhoven}},\ }\bibfield  {title} {\bibinfo {title} {{Perspective on
  Majorana bound-states in hybrid superconductor-semiconductor nanowires}},\
  }\href@noop {} {\bibfield  {journal} {\bibinfo  {journal} {arXiv:
  2406.17568}\ } (\bibinfo {year} {2024})}\BibitemShut {NoStop}%
\bibitem [{\citenamefont {Mourik}\ \emph {et~al.}(2012)\citenamefont {Mourik},
  \citenamefont {Zuo}, \citenamefont {Frolov}, \citenamefont {Plissard},
  \citenamefont {Bakkers},\ and\ \citenamefont {Kouwenhoven}}]{Mourik}%
  \BibitemOpen
  \bibfield  {author} {\bibinfo {author} {\bibfnamefont {V.}~\bibnamefont
  {Mourik}}, \bibinfo {author} {\bibfnamefont {K.}~\bibnamefont {Zuo}},
  \bibinfo {author} {\bibfnamefont {S.~M.}\ \bibnamefont {Frolov}}, \bibinfo
  {author} {\bibfnamefont {S.}~\bibnamefont {Plissard}}, \bibinfo {author}
  {\bibfnamefont {E.~P.}\ \bibnamefont {Bakkers}},\ and\ \bibinfo {author}
  {\bibfnamefont {L.~P.}\ \bibnamefont {Kouwenhoven}},\ }\bibfield  {title}
  {\bibinfo {title} {Signatures of {M}ajorana fermions in hybrid
  superconductor-semiconductor nanowire devices},\ }\href
  {https://doi.org/10.1126/science.1222360} {\bibfield  {journal} {\bibinfo
  {journal} {Science}\ }\textbf {\bibinfo {volume} {336}},\ \bibinfo {pages}
  {1003} (\bibinfo {year} {2012})}\BibitemShut {NoStop}%
\bibitem [{\citenamefont {Deng}\ \emph {et~al.}(2016)\citenamefont {Deng},
  \citenamefont {Vaitiek{\.e}nas}, \citenamefont {Hansen}, \citenamefont
  {Danon}, \citenamefont {Leijnse}, \citenamefont {Flensberg}, \citenamefont
  {Nyg{\aa}rd}, \citenamefont {Krogstrup},\ and\ \citenamefont
  {Marcus}}]{Deng2016}%
  \BibitemOpen
  \bibfield  {author} {\bibinfo {author} {\bibfnamefont {M.}~\bibnamefont
  {Deng}}, \bibinfo {author} {\bibfnamefont {S.}~\bibnamefont
  {Vaitiek{\.e}nas}}, \bibinfo {author} {\bibfnamefont {E.~B.}\ \bibnamefont
  {Hansen}}, \bibinfo {author} {\bibfnamefont {J.}~\bibnamefont {Danon}},
  \bibinfo {author} {\bibfnamefont {M.}~\bibnamefont {Leijnse}}, \bibinfo
  {author} {\bibfnamefont {K.}~\bibnamefont {Flensberg}}, \bibinfo {author}
  {\bibfnamefont {J.}~\bibnamefont {Nyg{\aa}rd}}, \bibinfo {author}
  {\bibfnamefont {P.}~\bibnamefont {Krogstrup}},\ and\ \bibinfo {author}
  {\bibfnamefont {C.~M.}\ \bibnamefont {Marcus}},\ }\bibfield  {title}
  {\bibinfo {title} {Majorana bound state in a coupled quantum-dot
  hybrid-nanowire system},\ }\href {https://doi.org/10.1126/science.aaf3961}
  {\bibfield  {journal} {\bibinfo  {journal} {Science}\ }\textbf {\bibinfo
  {volume} {354}},\ \bibinfo {pages} {1557} (\bibinfo {year}
  {2016})}\BibitemShut {NoStop}%
\bibitem [{\citenamefont {G{\"u}l}\ \emph {et~al.}(2018)\citenamefont
  {G{\"u}l}, \citenamefont {Zhang}, \citenamefont {Bommer}, \citenamefont
  {de~Moor}, \citenamefont {Car}, \citenamefont {Plissard}, \citenamefont
  {Bakkers}, \citenamefont {Geresdi}, \citenamefont {Watanabe}, \citenamefont
  {Taniguchi} \emph {et~al.}}]{Gul2018}%
  \BibitemOpen
  \bibfield  {author} {\bibinfo {author} {\bibfnamefont {{\"O}.}~\bibnamefont
  {G{\"u}l}}, \bibinfo {author} {\bibfnamefont {H.}~\bibnamefont {Zhang}},
  \bibinfo {author} {\bibfnamefont {J.~D.}\ \bibnamefont {Bommer}}, \bibinfo
  {author} {\bibfnamefont {M.~W.}\ \bibnamefont {de~Moor}}, \bibinfo {author}
  {\bibfnamefont {D.}~\bibnamefont {Car}}, \bibinfo {author} {\bibfnamefont
  {S.~R.}\ \bibnamefont {Plissard}}, \bibinfo {author} {\bibfnamefont {E.~P.}\
  \bibnamefont {Bakkers}}, \bibinfo {author} {\bibfnamefont {A.}~\bibnamefont
  {Geresdi}}, \bibinfo {author} {\bibfnamefont {K.}~\bibnamefont {Watanabe}},
  \bibinfo {author} {\bibfnamefont {T.}~\bibnamefont {Taniguchi}}, \emph
  {et~al.},\ }\bibfield  {title} {\bibinfo {title} {Ballistic {M}ajorana
  nanowire devices},\ }\href {https://doi.org/10.1038/s41565-017-0032-8}
  {\bibfield  {journal} {\bibinfo  {journal} {Nature Nanotechnology}\ }\textbf
  {\bibinfo {volume} {13}},\ \bibinfo {pages} {192} (\bibinfo {year}
  {2018})}\BibitemShut {NoStop}%
\bibitem [{\citenamefont {Song}\ \emph {et~al.}(2022)\citenamefont {Song},
  \citenamefont {Zhang}, \citenamefont {Pan}, \citenamefont {Liu},
  \citenamefont {Wang}, \citenamefont {Cao}, \citenamefont {Liu}, \citenamefont
  {Wen}, \citenamefont {Liao}, \citenamefont {Zhuo} \emph {et~al.}}]{Song2022}%
  \BibitemOpen
  \bibfield  {author} {\bibinfo {author} {\bibfnamefont {H.}~\bibnamefont
  {Song}}, \bibinfo {author} {\bibfnamefont {Z.}~\bibnamefont {Zhang}},
  \bibinfo {author} {\bibfnamefont {D.}~\bibnamefont {Pan}}, \bibinfo {author}
  {\bibfnamefont {D.}~\bibnamefont {Liu}}, \bibinfo {author} {\bibfnamefont
  {Z.}~\bibnamefont {Wang}}, \bibinfo {author} {\bibfnamefont {Z.}~\bibnamefont
  {Cao}}, \bibinfo {author} {\bibfnamefont {L.}~\bibnamefont {Liu}}, \bibinfo
  {author} {\bibfnamefont {L.}~\bibnamefont {Wen}}, \bibinfo {author}
  {\bibfnamefont {D.}~\bibnamefont {Liao}}, \bibinfo {author} {\bibfnamefont
  {R.}~\bibnamefont {Zhuo}}, \emph {et~al.},\ }\bibfield  {title} {\bibinfo
  {title} {Large zero bias peaks and dips in a four-terminal thin {I}n{A}s-{A}l
  nanowire device},\ }\href {https://doi.org/10.1103/PhysRevResearch.4.033235}
  {\bibfield  {journal} {\bibinfo  {journal} {Phys. Rev. Research}\ }\textbf
  {\bibinfo {volume} {4}},\ \bibinfo {pages} {033235} (\bibinfo {year}
  {2022})}\BibitemShut {NoStop}%
\bibitem [{\citenamefont {Wang}\ \emph
  {et~al.}(2022{\natexlab{a}})\citenamefont {Wang}, \citenamefont {Song},
  \citenamefont {Pan}, \citenamefont {Zhang}, \citenamefont {Miao},
  \citenamefont {Li}, \citenamefont {Cao}, \citenamefont {Zhang}, \citenamefont
  {Liu}, \citenamefont {Wen} \emph {et~al.}}]{WangZhaoyu}%
  \BibitemOpen
  \bibfield  {author} {\bibinfo {author} {\bibfnamefont {Z.}~\bibnamefont
  {Wang}}, \bibinfo {author} {\bibfnamefont {H.}~\bibnamefont {Song}}, \bibinfo
  {author} {\bibfnamefont {D.}~\bibnamefont {Pan}}, \bibinfo {author}
  {\bibfnamefont {Z.}~\bibnamefont {Zhang}}, \bibinfo {author} {\bibfnamefont
  {W.}~\bibnamefont {Miao}}, \bibinfo {author} {\bibfnamefont {R.}~\bibnamefont
  {Li}}, \bibinfo {author} {\bibfnamefont {Z.}~\bibnamefont {Cao}}, \bibinfo
  {author} {\bibfnamefont {G.}~\bibnamefont {Zhang}}, \bibinfo {author}
  {\bibfnamefont {L.}~\bibnamefont {Liu}}, \bibinfo {author} {\bibfnamefont
  {L.}~\bibnamefont {Wen}}, \emph {et~al.},\ }\bibfield  {title} {\bibinfo
  {title} {Plateau regions for zero-bias peaks within 5$\%$ of the quantized
  conductance value $2{e}^{2}/h$},\ }\href
  {https://doi.org/10.1103/PhysRevLett.129.167702} {\bibfield  {journal}
  {\bibinfo  {journal} {Phys. Rev. Lett.}\ }\textbf {\bibinfo {volume} {129}},\
  \bibinfo {pages} {167702} (\bibinfo {year} {2022}{\natexlab{a}})}\BibitemShut
  {NoStop}%
\bibitem [{\citenamefont {Aghaee}\ \emph {et~al.}(2023)\citenamefont {Aghaee},
  \citenamefont {Akkala}, \citenamefont {Alam}, \citenamefont {Ali},
  \citenamefont {Alcaraz~Ramirez}, \citenamefont {Andrzejczuk}, \citenamefont
  {Antipov}, \citenamefont {Aseev}, \citenamefont {Astafev}, \citenamefont
  {Bauer} \emph {et~al.}}]{MS_2023}%
  \BibitemOpen
  \bibfield  {author} {\bibinfo {author} {\bibfnamefont {M.}~\bibnamefont
  {Aghaee}}, \bibinfo {author} {\bibfnamefont {A.}~\bibnamefont {Akkala}},
  \bibinfo {author} {\bibfnamefont {Z.}~\bibnamefont {Alam}}, \bibinfo {author}
  {\bibfnamefont {R.}~\bibnamefont {Ali}}, \bibinfo {author} {\bibfnamefont
  {A.}~\bibnamefont {Alcaraz~Ramirez}}, \bibinfo {author} {\bibfnamefont
  {M.}~\bibnamefont {Andrzejczuk}}, \bibinfo {author} {\bibfnamefont {A.~E.}\
  \bibnamefont {Antipov}}, \bibinfo {author} {\bibfnamefont {P.}~\bibnamefont
  {Aseev}}, \bibinfo {author} {\bibfnamefont {M.}~\bibnamefont {Astafev}},
  \bibinfo {author} {\bibfnamefont {B.}~\bibnamefont {Bauer}}, \emph {et~al.}
  (\bibinfo {collaboration} {Microsoft Quantum}),\ }\bibfield  {title}
  {\bibinfo {title} {{InAs-Al} hybrid devices passing the topological gap
  protocol},\ }\href {https://doi.org/10.1103/PhysRevB.107.245423} {\bibfield
  {journal} {\bibinfo  {journal} {Phys. Rev. B}\ }\textbf {\bibinfo {volume}
  {107}},\ \bibinfo {pages} {245423} (\bibinfo {year} {2023})}\BibitemShut
  {NoStop}%
\bibitem [{\citenamefont {Dvir}\ \emph {et~al.}(2023)\citenamefont {Dvir},
  \citenamefont {Wang}, \citenamefont {van Loo}, \citenamefont {Liu},
  \citenamefont {Mazur}, \citenamefont {Bordin}, \citenamefont {Haaf},
  \citenamefont {Wang}, \citenamefont {Driel}, \citenamefont {Zatelli} \emph
  {et~al.}}]{Delft_Kitaev}%
  \BibitemOpen
  \bibfield  {author} {\bibinfo {author} {\bibfnamefont {T.}~\bibnamefont
  {Dvir}}, \bibinfo {author} {\bibfnamefont {G.}~\bibnamefont {Wang}}, \bibinfo
  {author} {\bibfnamefont {N.}~\bibnamefont {van Loo}}, \bibinfo {author}
  {\bibfnamefont {C.-X.}\ \bibnamefont {Liu}}, \bibinfo {author} {\bibfnamefont
  {G.}~\bibnamefont {Mazur}}, \bibinfo {author} {\bibfnamefont
  {A.}~\bibnamefont {Bordin}}, \bibinfo {author} {\bibfnamefont
  {S.}~\bibnamefont {Haaf}}, \bibinfo {author} {\bibfnamefont {J.-Y.}\
  \bibnamefont {Wang}}, \bibinfo {author} {\bibfnamefont {D.}~\bibnamefont
  {Driel}}, \bibinfo {author} {\bibfnamefont {F.}~\bibnamefont {Zatelli}},
  \emph {et~al.},\ }\bibfield  {title} {\bibinfo {title} {Realization of a
  minimal {Kitaev} chain in coupled quantum dots},\ }\href
  {https://doi.org/10.1038/s41586-022-05585-1} {\bibfield  {journal} {\bibinfo
  {journal} {Nature}\ }\textbf {\bibinfo {volume} {614}},\ \bibinfo {pages}
  {445} (\bibinfo {year} {2023})}\BibitemShut {NoStop}%
\bibitem [{\citenamefont {Albrecht}\ \emph {et~al.}(2016)\citenamefont
  {Albrecht}, \citenamefont {Higginbotham}, \citenamefont {Madsen},
  \citenamefont {Kuemmeth}, \citenamefont {Jespersen}, \citenamefont {Nygård},
  \citenamefont {Krogstrup},\ and\ \citenamefont {Marcus}}]{Albrecht}%
  \BibitemOpen
  \bibfield  {author} {\bibinfo {author} {\bibfnamefont {S.~M.}\ \bibnamefont
  {Albrecht}}, \bibinfo {author} {\bibfnamefont {A.~P.}\ \bibnamefont
  {Higginbotham}}, \bibinfo {author} {\bibfnamefont {M.}~\bibnamefont
  {Madsen}}, \bibinfo {author} {\bibfnamefont {F.}~\bibnamefont {Kuemmeth}},
  \bibinfo {author} {\bibfnamefont {T.~S.}\ \bibnamefont {Jespersen}}, \bibinfo
  {author} {\bibfnamefont {J.}~\bibnamefont {Nygård}}, \bibinfo {author}
  {\bibfnamefont {P.}~\bibnamefont {Krogstrup}},\ and\ \bibinfo {author}
  {\bibfnamefont {C.~M.}\ \bibnamefont {Marcus}},\ }\bibfield  {title}
  {\bibinfo {title} {Exponential protection of zero modes in {M}ajorana
  islands},\ }\href {https://doi.org/10.1038/nature17162} {\bibfield  {journal}
  {\bibinfo  {journal} {Nature}\ }\textbf {\bibinfo {volume} {531}},\ \bibinfo
  {pages} {206} (\bibinfo {year} {2016})}\BibitemShut {NoStop}%
\bibitem [{\citenamefont {Pan}\ \emph {et~al.}(2022)\citenamefont {Pan},
  \citenamefont {Song}, \citenamefont {Zhang}, \citenamefont {Liu},
  \citenamefont {Wen}, \citenamefont {Liao}, \citenamefont {Zhuo},
  \citenamefont {Wang}, \citenamefont {Zhang}, \citenamefont {Yang} \emph
  {et~al.}}]{PanCPL}%
  \BibitemOpen
  \bibfield  {author} {\bibinfo {author} {\bibfnamefont {D.}~\bibnamefont
  {Pan}}, \bibinfo {author} {\bibfnamefont {H.}~\bibnamefont {Song}}, \bibinfo
  {author} {\bibfnamefont {S.}~\bibnamefont {Zhang}}, \bibinfo {author}
  {\bibfnamefont {L.}~\bibnamefont {Liu}}, \bibinfo {author} {\bibfnamefont
  {L.}~\bibnamefont {Wen}}, \bibinfo {author} {\bibfnamefont {D.}~\bibnamefont
  {Liao}}, \bibinfo {author} {\bibfnamefont {R.}~\bibnamefont {Zhuo}}, \bibinfo
  {author} {\bibfnamefont {Z.}~\bibnamefont {Wang}}, \bibinfo {author}
  {\bibfnamefont {Z.}~\bibnamefont {Zhang}}, \bibinfo {author} {\bibfnamefont
  {S.}~\bibnamefont {Yang}}, \emph {et~al.},\ }\bibfield  {title} {\bibinfo
  {title} {In situ epitaxy of pure phase ultra-thin {InAs-Al} nanowires for
  quantum devices},\ }\href {https://doi.org/10.1088/0256-307X/39/5/058101}
  {\bibfield  {journal} {\bibinfo  {journal} {Chinese Physics Letters}\
  }\textbf {\bibinfo {volume} {39}},\ \bibinfo {eid} {058101} (\bibinfo {year}
  {2022})}\BibitemShut {NoStop}%
\bibitem [{\citenamefont {Lee}\ \emph {et~al.}(2014)\citenamefont {Lee},
  \citenamefont {Jiang}, \citenamefont {Houzet}, \citenamefont {Aguado},
  \citenamefont {Lieber},\ and\ \citenamefont {De~Franceschi}}]{Silvano2014}%
  \BibitemOpen
  \bibfield  {author} {\bibinfo {author} {\bibfnamefont {E.~J.}\ \bibnamefont
  {Lee}}, \bibinfo {author} {\bibfnamefont {X.}~\bibnamefont {Jiang}}, \bibinfo
  {author} {\bibfnamefont {M.}~\bibnamefont {Houzet}}, \bibinfo {author}
  {\bibfnamefont {R.}~\bibnamefont {Aguado}}, \bibinfo {author} {\bibfnamefont
  {C.~M.}\ \bibnamefont {Lieber}},\ and\ \bibinfo {author} {\bibfnamefont
  {S.}~\bibnamefont {De~Franceschi}},\ }\bibfield  {title} {\bibinfo {title}
  {Spin-resolved {A}ndreev levels and parity crossings in hybrid
  superconductor--semiconductor nanostructures},\ }\href
  {https://doi.org/10.1038/nnano.2013.267} {\bibfield  {journal} {\bibinfo
  {journal} {Nature Nanotechnology}\ }\textbf {\bibinfo {volume} {9}},\
  \bibinfo {pages} {79} (\bibinfo {year} {2014})}\BibitemShut {NoStop}%
\bibitem [{\citenamefont {Aleiner}\ and\ \citenamefont
  {Glazman}(1998)}]{Glazman_CB}%
  \BibitemOpen
  \bibfield  {author} {\bibinfo {author} {\bibfnamefont {I.~L.}\ \bibnamefont
  {Aleiner}}\ and\ \bibinfo {author} {\bibfnamefont {L.~I.}\ \bibnamefont
  {Glazman}},\ }\bibfield  {title} {\bibinfo {title} {Mesoscopic charge
  quantization},\ }\href {https://doi.org/10.1103/PhysRevB.57.9608} {\bibfield
  {journal} {\bibinfo  {journal} {Phys. Rev. B}\ }\textbf {\bibinfo {volume}
  {57}},\ \bibinfo {pages} {9608} (\bibinfo {year} {1998})}\BibitemShut
  {NoStop}%
\bibitem [{\citenamefont {Albrecht}\ \emph {et~al.}(2017)\citenamefont
  {Albrecht}, \citenamefont {Hansen}, \citenamefont {Higginbotham},
  \citenamefont {Kuemmeth}, \citenamefont {Jespersen}, \citenamefont {Nygård},
  \citenamefont {Krogstrup}, \citenamefont {Danon}, \citenamefont {Flensberg},\
  and\ \citenamefont {Marcus}}]{Albrecht_PRL}%
  \BibitemOpen
  \bibfield  {author} {\bibinfo {author} {\bibfnamefont {S.~M.}\ \bibnamefont
  {Albrecht}}, \bibinfo {author} {\bibfnamefont {E.~B.}\ \bibnamefont
  {Hansen}}, \bibinfo {author} {\bibfnamefont {A.~P.}\ \bibnamefont
  {Higginbotham}}, \bibinfo {author} {\bibfnamefont {F.}~\bibnamefont
  {Kuemmeth}}, \bibinfo {author} {\bibfnamefont {T.~S.}\ \bibnamefont
  {Jespersen}}, \bibinfo {author} {\bibfnamefont {J.}~\bibnamefont {Nygård}},
  \bibinfo {author} {\bibfnamefont {P.}~\bibnamefont {Krogstrup}}, \bibinfo
  {author} {\bibfnamefont {J.}~\bibnamefont {Danon}}, \bibinfo {author}
  {\bibfnamefont {K.}~\bibnamefont {Flensberg}},\ and\ \bibinfo {author}
  {\bibfnamefont {C.~M.}\ \bibnamefont {Marcus}},\ }\bibfield  {title}
  {\bibinfo {title} {Transport signatures of quasiparticle poisoning in a
  majorana island},\ }\href {https://doi.org/10.1103/PhysRevLett.118.137701}
  {\bibfield  {journal} {\bibinfo  {journal} {Physical Review Letters}\
  }\textbf {\bibinfo {volume} {118}},\ \bibinfo {pages} {137701} (\bibinfo
  {year} {2017})}\BibitemShut {NoStop}%
\bibitem [{\citenamefont {Zhang}\ \emph {et~al.}(2022)\citenamefont {Zhang},
  \citenamefont {Wang}, \citenamefont {Pan}, \citenamefont {Li}, \citenamefont
  {Lu}, \citenamefont {Li}, \citenamefont {Zhang}, \citenamefont {Liu},
  \citenamefont {Cao}, \citenamefont {Liu}, \citenamefont {Wen}, \citenamefont
  {Liao}, \citenamefont {Zhuo}, \citenamefont {Shang}, \citenamefont {Liu},
  \citenamefont {Zhao},\ and\ \citenamefont {Zhang}}]{ZhangShan}%
  \BibitemOpen
  \bibfield  {author} {\bibinfo {author} {\bibfnamefont {S.}~\bibnamefont
  {Zhang}}, \bibinfo {author} {\bibfnamefont {Z.}~\bibnamefont {Wang}},
  \bibinfo {author} {\bibfnamefont {D.}~\bibnamefont {Pan}}, \bibinfo {author}
  {\bibfnamefont {H.}~\bibnamefont {Li}}, \bibinfo {author} {\bibfnamefont
  {S.}~\bibnamefont {Lu}}, \bibinfo {author} {\bibfnamefont {Z.}~\bibnamefont
  {Li}}, \bibinfo {author} {\bibfnamefont {G.}~\bibnamefont {Zhang}}, \bibinfo
  {author} {\bibfnamefont {D.}~\bibnamefont {Liu}}, \bibinfo {author}
  {\bibfnamefont {Z.}~\bibnamefont {Cao}}, \bibinfo {author} {\bibfnamefont
  {L.}~\bibnamefont {Liu}}, \bibinfo {author} {\bibfnamefont {L.}~\bibnamefont
  {Wen}}, \bibinfo {author} {\bibfnamefont {D.}~\bibnamefont {Liao}}, \bibinfo
  {author} {\bibfnamefont {R.}~\bibnamefont {Zhuo}}, \bibinfo {author}
  {\bibfnamefont {R.}~\bibnamefont {Shang}}, \bibinfo {author} {\bibfnamefont
  {D.~E.}\ \bibnamefont {Liu}}, \bibinfo {author} {\bibfnamefont
  {J.}~\bibnamefont {Zhao}},\ and\ \bibinfo {author} {\bibfnamefont
  {H.}~\bibnamefont {Zhang}},\ }\bibfield  {title} {\bibinfo {title}
  {Suppressing andreev bound state zero bias peaks using a strongly dissipative
  lead},\ }\href {https://doi.org/10.1103/PhysRevLett.128.076803} {\bibfield
  {journal} {\bibinfo  {journal} {Phys. Rev. Lett.}\ }\textbf {\bibinfo
  {volume} {128}},\ \bibinfo {pages} {076803} (\bibinfo {year}
  {2022})}\BibitemShut {NoStop}%
\bibitem [{\citenamefont {Wang}\ \emph
  {et~al.}(2022{\natexlab{b}})\citenamefont {Wang}, \citenamefont {Zhang},
  \citenamefont {Pan}, \citenamefont {Zhang}, \citenamefont {Xia},
  \citenamefont {Li}, \citenamefont {Liu}, \citenamefont {Cao}, \citenamefont
  {Liu}, \citenamefont {Wen}, \citenamefont {Liao}, \citenamefont {Zhuo},
  \citenamefont {Li}, \citenamefont {Liu}, \citenamefont {Shang}, \citenamefont
  {Zhao},\ and\ \citenamefont {Zhang}}]{WangZhichuan}%
  \BibitemOpen
  \bibfield  {author} {\bibinfo {author} {\bibfnamefont {Z.}~\bibnamefont
  {Wang}}, \bibinfo {author} {\bibfnamefont {S.}~\bibnamefont {Zhang}},
  \bibinfo {author} {\bibfnamefont {D.}~\bibnamefont {Pan}}, \bibinfo {author}
  {\bibfnamefont {G.}~\bibnamefont {Zhang}}, \bibinfo {author} {\bibfnamefont
  {Z.}~\bibnamefont {Xia}}, \bibinfo {author} {\bibfnamefont {Z.}~\bibnamefont
  {Li}}, \bibinfo {author} {\bibfnamefont {D.}~\bibnamefont {Liu}}, \bibinfo
  {author} {\bibfnamefont {Z.}~\bibnamefont {Cao}}, \bibinfo {author}
  {\bibfnamefont {L.}~\bibnamefont {Liu}}, \bibinfo {author} {\bibfnamefont
  {L.}~\bibnamefont {Wen}}, \bibinfo {author} {\bibfnamefont {D.}~\bibnamefont
  {Liao}}, \bibinfo {author} {\bibfnamefont {R.}~\bibnamefont {Zhuo}}, \bibinfo
  {author} {\bibfnamefont {Y.}~\bibnamefont {Li}}, \bibinfo {author}
  {\bibfnamefont {D.~E.}\ \bibnamefont {Liu}}, \bibinfo {author} {\bibfnamefont
  {R.}~\bibnamefont {Shang}}, \bibinfo {author} {\bibfnamefont
  {J.}~\bibnamefont {Zhao}},\ and\ \bibinfo {author} {\bibfnamefont
  {H.}~\bibnamefont {Zhang}},\ }\bibfield  {title} {\bibinfo {title} {Large
  {A}ndreev bound state zero-bias peaks in a weakly dissipative environment},\
  }\href {https://doi.org/10.1103/PhysRevB.106.205421} {\bibfield  {journal}
  {\bibinfo  {journal} {Phys. Rev. B}\ }\textbf {\bibinfo {volume} {106}},\
  \bibinfo {pages} {205421} (\bibinfo {year} {2022}{\natexlab{b}})}\BibitemShut
  {NoStop}%
\bibitem [{\citenamefont {Valentini}\ \emph {et~al.}(2022)\citenamefont
  {Valentini}, \citenamefont {Borovkov}, \citenamefont {Prada}, \citenamefont
  {Martí-Sánchez}, \citenamefont {Botifoll}, \citenamefont {Hofmann},
  \citenamefont {Arbiol}, \citenamefont {Aguado}, \citenamefont {San-Jose},\
  and\ \citenamefont {Katsaros}}]{Valentini_Nature}%
  \BibitemOpen
  \bibfield  {author} {\bibinfo {author} {\bibfnamefont {M.}~\bibnamefont
  {Valentini}}, \bibinfo {author} {\bibfnamefont {M.}~\bibnamefont {Borovkov}},
  \bibinfo {author} {\bibfnamefont {E.}~\bibnamefont {Prada}}, \bibinfo
  {author} {\bibfnamefont {S.}~\bibnamefont {Martí-Sánchez}}, \bibinfo
  {author} {\bibfnamefont {M.}~\bibnamefont {Botifoll}}, \bibinfo {author}
  {\bibfnamefont {A.}~\bibnamefont {Hofmann}}, \bibinfo {author} {\bibfnamefont
  {J.}~\bibnamefont {Arbiol}}, \bibinfo {author} {\bibfnamefont
  {R.}~\bibnamefont {Aguado}}, \bibinfo {author} {\bibfnamefont
  {P.}~\bibnamefont {San-Jose}},\ and\ \bibinfo {author} {\bibfnamefont
  {G.}~\bibnamefont {Katsaros}},\ }\bibfield  {title} {\bibinfo {title}
  {Majorana-like coulomb spectroscopy in the absence of zero-bias peaks},\
  }\href {https://doi.org/10.1038/s41586-022-05382-w} {\bibfield  {journal}
  {\bibinfo  {journal} {Nature}\ }\textbf {\bibinfo {volume} {612}},\ \bibinfo
  {pages} {442} (\bibinfo {year} {2022})}\BibitemShut {NoStop}%
\bibitem [{osc(2024)}]{oscillation_Zenodo}%
  \BibitemOpen
  \bibfield  {title} {\bibinfo {title} {10.5281/zenodo.13899821},\ }\href
  {https://doi.org/10.5281/zenodo.13899821} {\bibfield  {journal} {\bibinfo
  {journal} {Zenodo}\ } (\bibinfo {year} {2024})}\BibitemShut {NoStop}%
\end{thebibliography}%


\begin{thebibliography}{0}%
\makeatletter
\providecommand \@ifxundefined [1]{%
 \@ifx{#1\undefined}
}%
\providecommand \@ifnum [1]{%
 \ifnum #1\expandafter \@firstoftwo
 \else \expandafter \@secondoftwo
 \fi
}%
\providecommand \@ifx [1]{%
 \ifx #1\expandafter \@firstoftwo
 \else \expandafter \@secondoftwo
 \fi
}%
\providecommand \natexlab [1]{#1}%
\providecommand \enquote  [1]{``#1''}%
\providecommand \bibnamefont  [1]{#1}%
\providecommand \bibfnamefont [1]{#1}%
\providecommand \citenamefont [1]{#1}%
\providecommand \href@noop [0]{\@secondoftwo}%
\providecommand \href [0]{\begingroup \@sanitize@url \@href}%
\providecommand \@href[1]{\@@startlink{#1}\@@href}%
\providecommand \@@href[1]{\endgroup#1\@@endlink}%
\providecommand \@sanitize@url [0]{\catcode `\\12\catcode `\$12\catcode
  `\&12\catcode `\#12\catcode `\^12\catcode `\_12\catcode `\%12\relax}%
\providecommand \@@startlink[1]{}%
\providecommand \@@endlink[0]{}%
\providecommand \url  [0]{\begingroup\@sanitize@url \@url }%
\providecommand \@url [1]{\endgroup\@href {#1}{\urlprefix }}%
\providecommand \urlprefix  [0]{URL }%
\providecommand \Eprint [0]{\href }%
\providecommand \doibase [0]{https://doi.org/}%
\providecommand \selectlanguage [0]{\@gobble}%
\providecommand \bibinfo  [0]{\@secondoftwo}%
\providecommand \bibfield  [0]{\@secondoftwo}%
\providecommand \translation [1]{[#1]}%
\providecommand \BibitemOpen [0]{}%
\providecommand \bibitemStop [0]{}%
\providecommand \bibitemNoStop [0]{.\EOS\space}%
\providecommand \EOS [0]{\spacefactor3000\relax}%
\providecommand \BibitemShut  [1]{\csname bibitem#1\endcsname}%
\let\auto@bib@innerbib\@empty
\end{thebibliography}%




\end{document}


\title{Supplementary Information for ``Coulomb blockade in open superconducting islands on InAs nanowires''}

\author{Huading Song}
\email{equal contribution}
\email{songhd@baqis.ac.cn}
\affiliation{Beijing Academy of Quantum Information Sciences, Beijing 100193, China}

\author{Zhaoyu Wang}
\email{equal contribution}
\affiliation{State Key Laboratory of Low Dimensional Quantum Physics, Department of Physics, Tsinghua University, Beijing 100084, China}

\author{Dong Pan}
\email{equal contribution}
\affiliation{State Key Laboratory of Superlattices and Microstructures, Institute of Semiconductors, Chinese Academy of Sciences, P.O. Box 912, Beijing 100083, China}

\author{Jiaye Xu}
\affiliation{State Key Laboratory of Low Dimensional Quantum Physics, Department of Physics, Tsinghua University, Beijing 100084, China}

\author{Yuqing Wang}
\affiliation{Beijing Academy of Quantum Information Sciences, Beijing 100193, China}

\author{Zhan Cao}
\affiliation{Beijing Academy of Quantum Information Sciences, Beijing 100193, China}

\author{Dong E. Liu}
\affiliation{State Key Laboratory of Low Dimensional Quantum Physics, Department of Physics, Tsinghua University, Beijing 100084, China}
\affiliation{Beijing Academy of Quantum Information Sciences, Beijing 100193, China}
\affiliation{Frontier Science Center for Quantum Information, Beijing 100084, China}
\affiliation{Hefei National Laboratory, Hefei 230088, China}

\author{Ke He}
\affiliation{State Key Laboratory of Low Dimensional Quantum Physics, Department of Physics, Tsinghua University, Beijing 100084, China}
\affiliation{Beijing Academy of Quantum Information Sciences, Beijing 100193, China}
\affiliation{Frontier Science Center for Quantum Information, Beijing 100084, China}
\affiliation{Hefei National Laboratory, Hefei 230088, China}

\author{Runan Shang}
\email{shangrn@baqis.ac.cn}
\affiliation{Beijing Academy of Quantum Information Sciences, Beijing 100193, China}
\affiliation{Hefei National Laboratory, Hefei 230088, China}

\author{Jianhua Zhao}
\affiliation{State Key Laboratory of Superlattices and Microstructures, Institute of Semiconductors, Chinese Academy of Sciences, P.O. Box 912, Beijing 100083, China}

\author{Hao Zhang}
\email{hzquantum@mail.tsinghua.edu.cn}
\affiliation{State Key Laboratory of Low Dimensional Quantum Physics, Department of Physics, Tsinghua University, Beijing 100084, China}
\affiliation{Beijing Academy of Quantum Information Sciences, Beijing 100193, China}
\affiliation{Frontier Science Center for Quantum Information, Beijing 100084, China}

\maketitle

\captionsetup[figure]{name={Fig. S},justification=raggedright,singlelinecheck=off}
\captionsetup[table]{name={Table. S},justification=raggedright,singlelinecheck=off}

\begin{figure}[htb]
\includegraphics[width=\columnwidth]{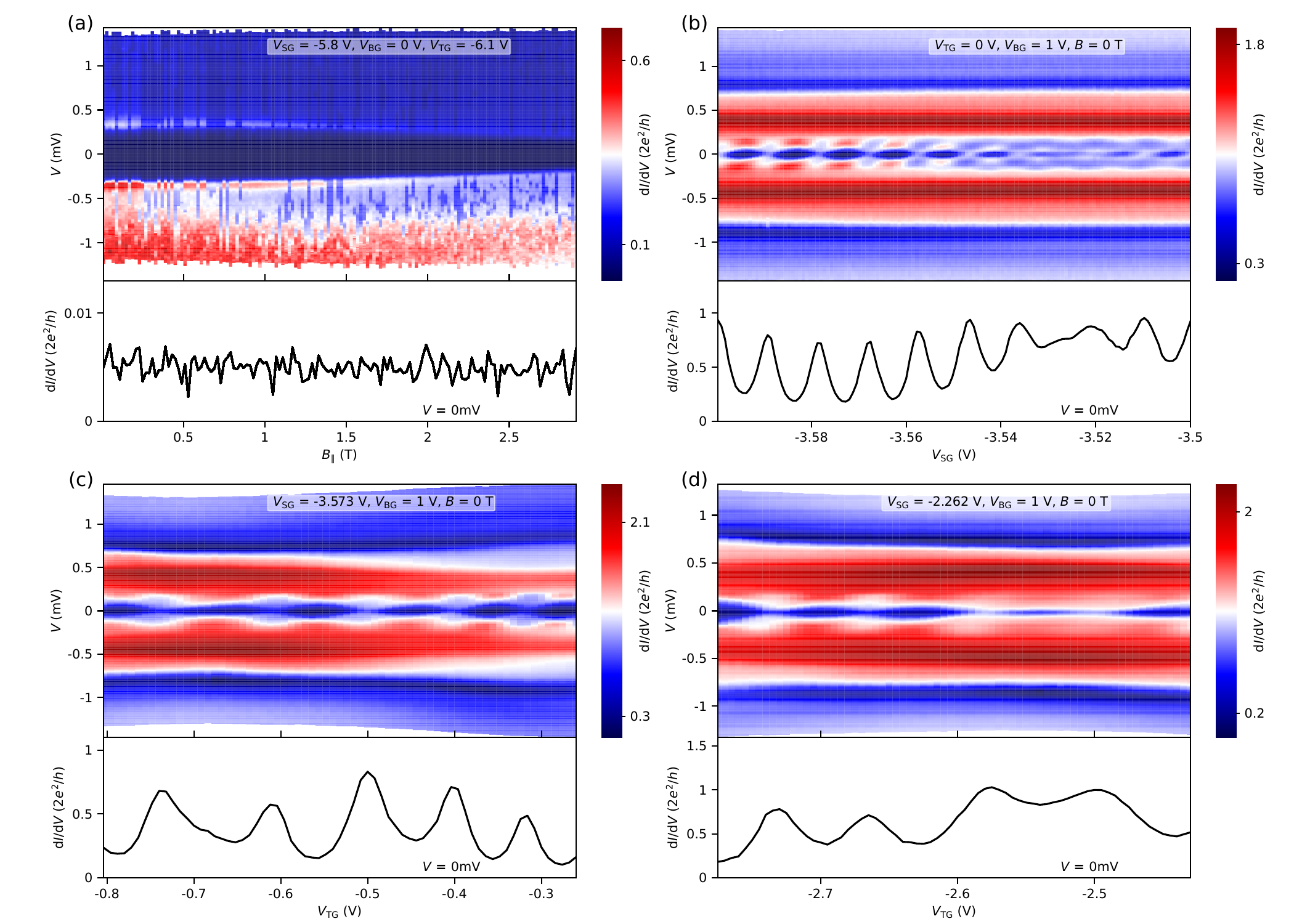}
\centering
\caption{Additional tunneling spectroscopy results of device A. (a) Hard gap tunneling spectroscopy of device A with parallel magnetic field, the critical field is \textgreater3T. (b) dI/dV as a function of $V_\mathrm{SG}$ and $V_\mathrm{bias}$ under B = 0 T, demonstrating clear Coulomb blockade effect. (c,d) dI/dV as a function of $V_\mathrm{TG}$ and $V_\mathrm{bias}$ under B= 0 T. $V_\mathrm{TG}$ also reveals the Coulomb blockade effect with much larger period due to crosstalk between $V_\mathrm{TG}$ and $V_\mathrm{SG}$.}
\label{sfig1}
\end{figure}

\begin{table}[htb]
\includegraphics[width=\columnwidth]{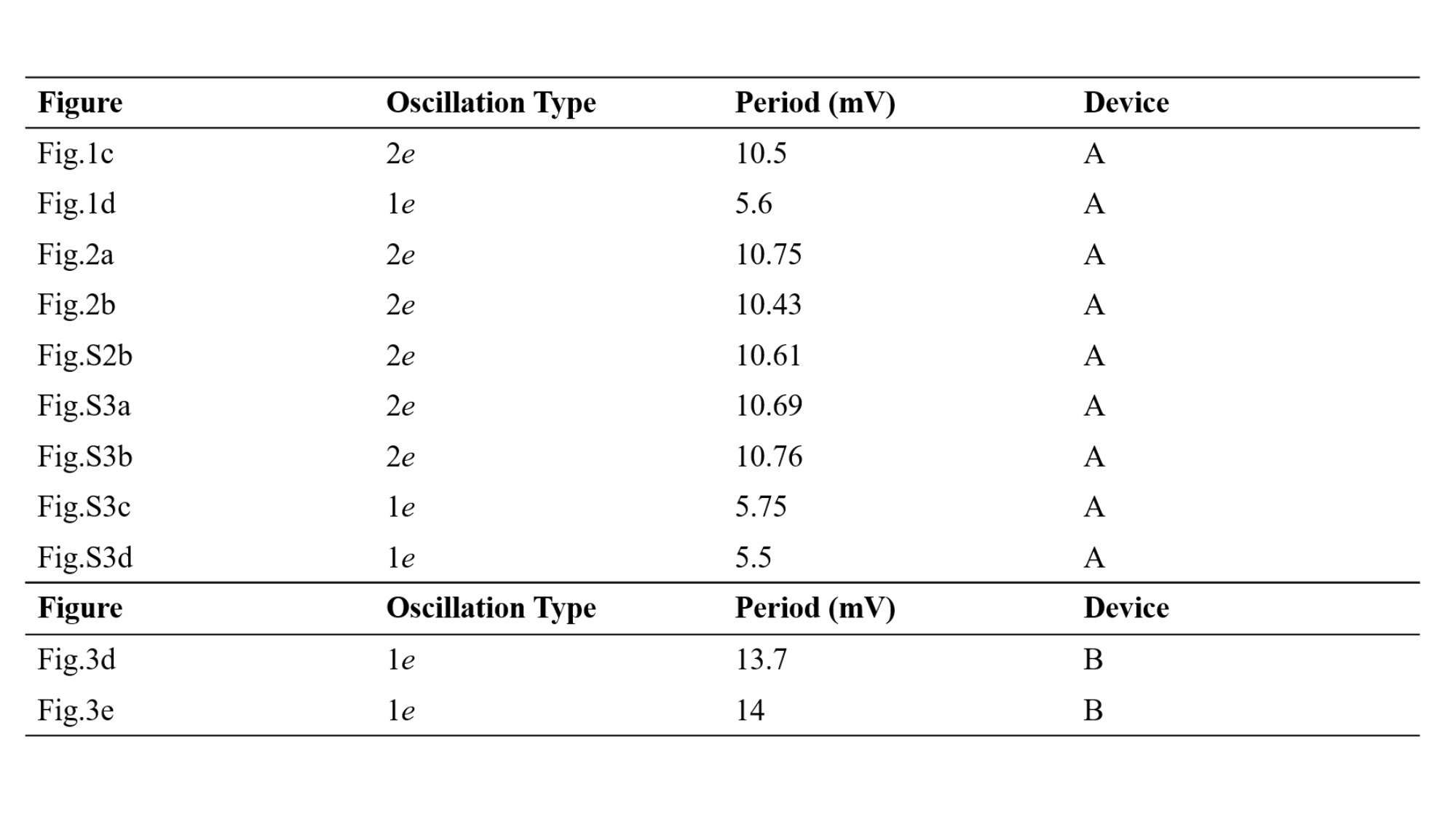}
\centering
\caption{Oscillation period of the Figures. The average oscillation period of device A is 10.62 mV for 2$e$, and 5.62 mV for 1$e$. The average oscillation period of 1$e$ oscillation of device B is 13.85 mV.}
\label{sTable1}
\end{table}

\begin{figure}[htb]
\includegraphics[width=\columnwidth]{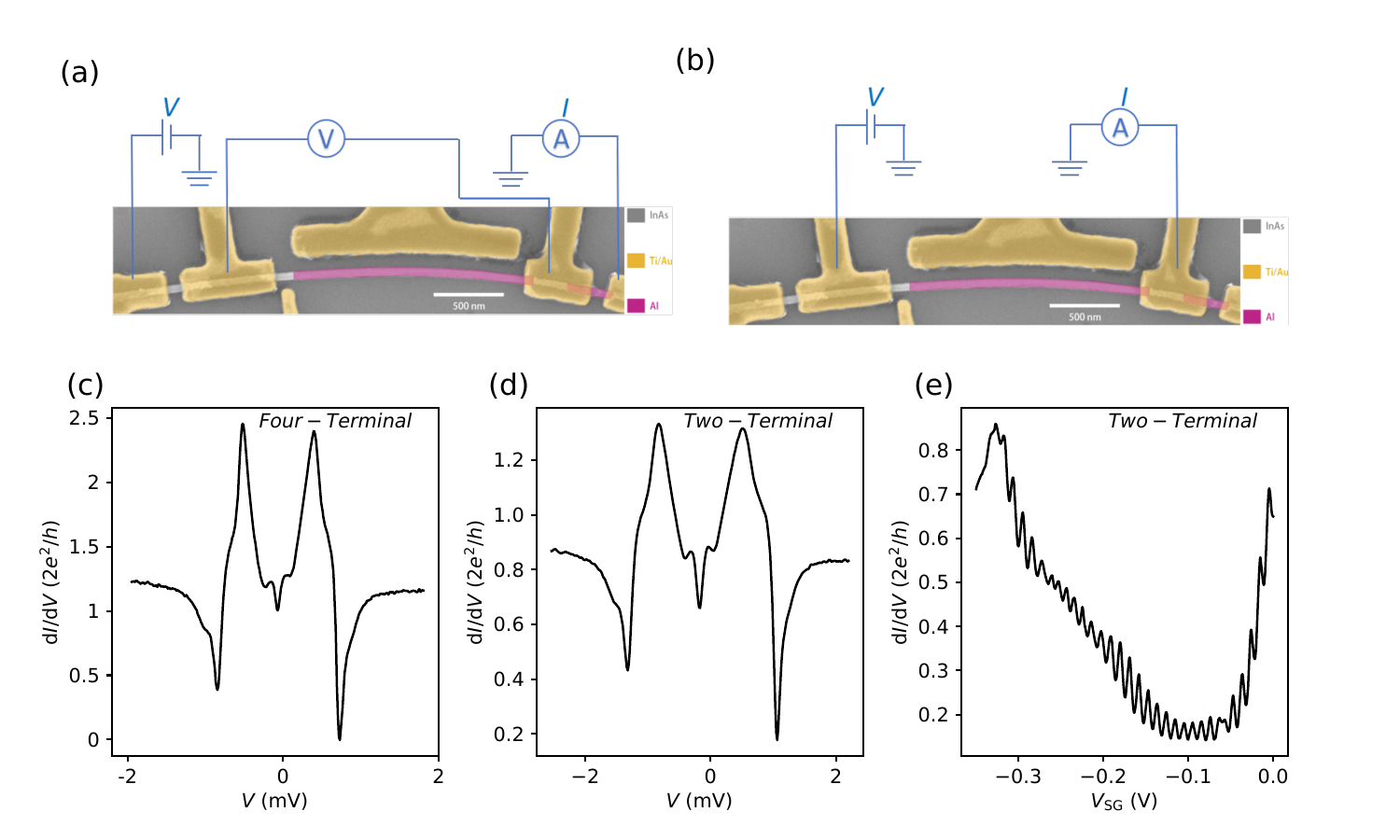}
\centering
\caption{Schematic diagram of four-terminal and two-terminal measurement circuits. The four-terminal setup can directly eliminate the additional influence of circuit and device contact resistances. (c,d) Four-terminal and two-terminal dI/dV results of device A with same gate parameters ($V_\mathrm{TG}=V_\mathrm{BG}=V_\mathrm{SG}=0 V$). The contact resistance is obtained by subtracting the results of four-terminal setup from the two-terminal setup in dI/dV saturation region ($V_\mathrm{bias} \gg \Delta$). After removing the calibrated circuit resistance (3.85 kOhm, mainly from RC filters), it was found that the contact resistance of the device is only $\sim$250 Ohm, this value matches the typical contact resistance of semiconductor-superconductor nanowire devices, indicating that the fabrication process produces good ohmic contacts.}
\label{sfig2}
\end{figure}

\begin{figure}[htb]
\includegraphics[width=\columnwidth]{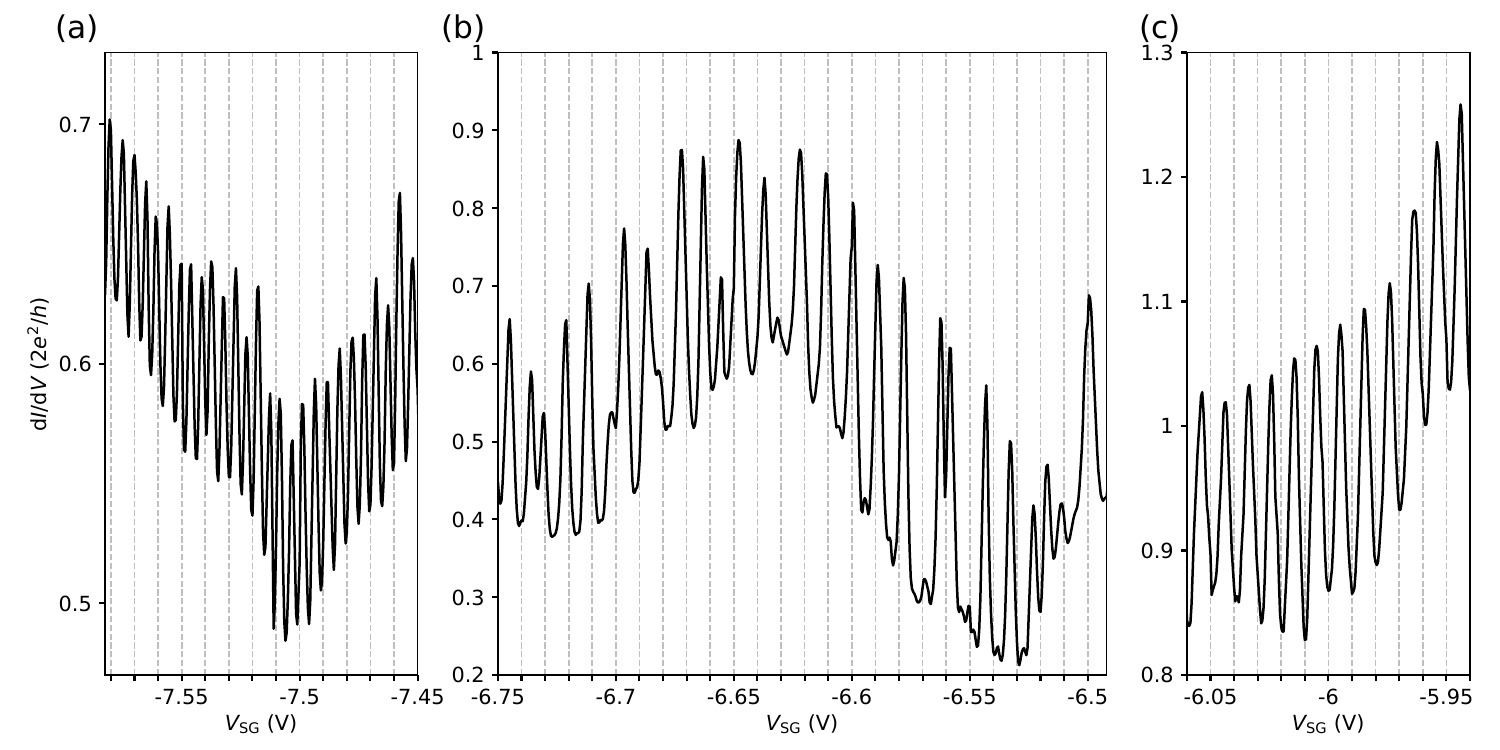}
\centering
\caption{2$e$ to 1$e$ period transition analysis. (a-c) dI/dV tuned by $V_\mathrm{SG}$ with $V_\mathrm{BG}= 1V$, $V_\mathrm{TG}= 8V$. (a) Uniform $1e$-period oscillation region. (b) The oscillation peak spacing becomes non-equal in the transition region between 2$e$- and $1e$-period. (c) Uniform $2e$-period oscillation region. }
\label{sfig3}
\end{figure}

\begin{figure}[htb]
\includegraphics[width=\columnwidth]{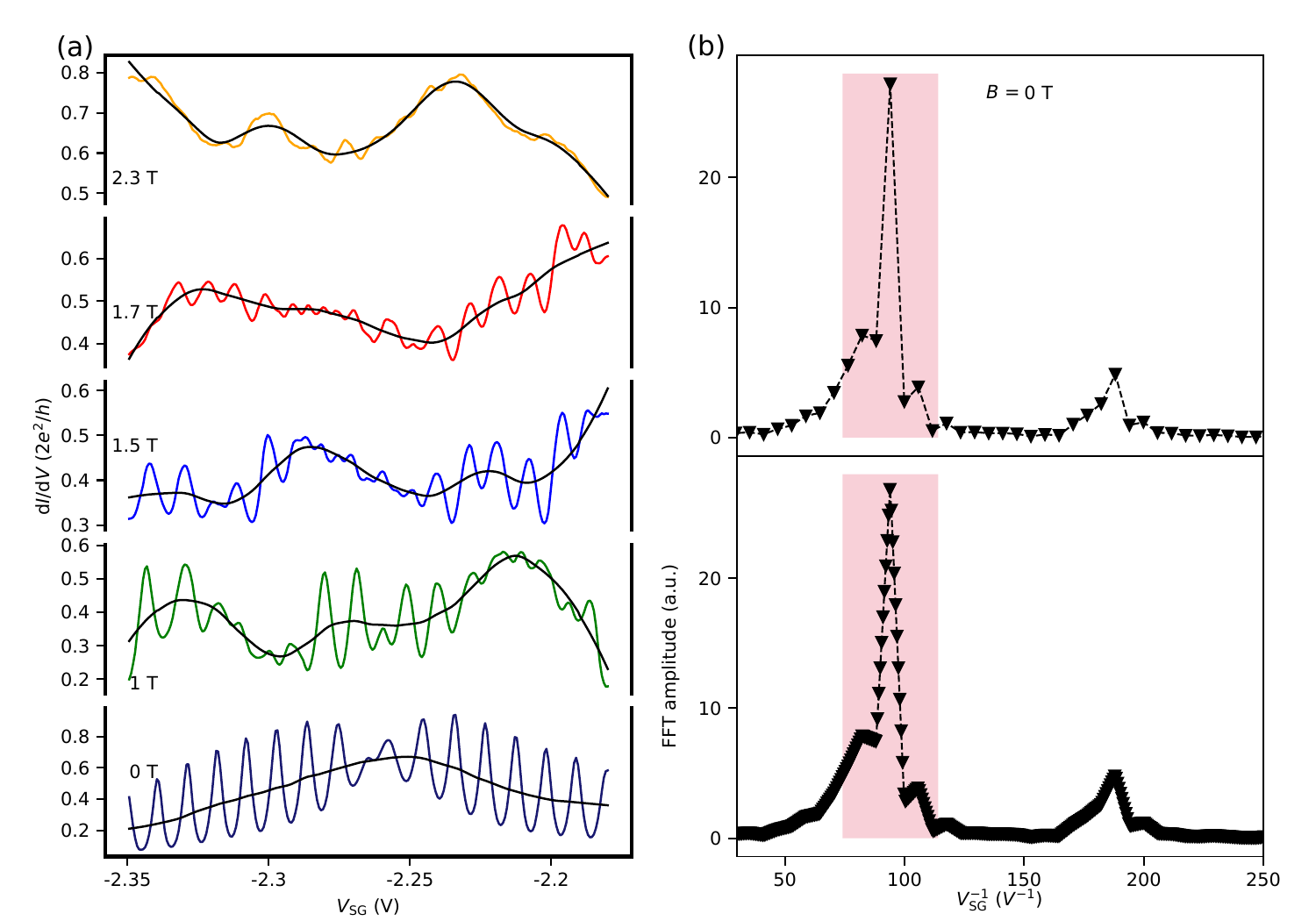}
\centering
\caption{Procedure to extract oscillation signal and oscillation FFT amplitudes. 1) The raw data are smoothed by using Savgol filter in Python environment. The smooth process contains two levels of filtering: The first level uses window length = 60, poly-order = 2. The second level uses window length = 30, poly-order = 2. The raw data and smoothed curves of Fig. 2(c) are shown in (a). 2) The oscillation signal $\Delta$G is calculated by removing the smoothed curve from raw data. 3) Performing fast Fourier transform (FFT) analysis on $\Delta$G [up panel of (b)]. 4) The FFT curve is interpolated into ten-times data points for better integration [bottom panel of (b)]. 5) The interpolated FFT spectrum is integrated within a frequency window of 40 $V^{-1}$ around the oscillation peak frequency [red shadowed region in (b)]. 6) The oscillation amplitude is determined as:
$$A_\mathrm{CB}= \int_{f_\mathrm{CB}-\frac{1}{2}W}^{f_\mathrm{CB}+\frac{1}{2}W} A(f)\, dx $$ 
Where $A_\mathrm{CB}$ is the oscillation amplitude, $f_\mathrm{CB}$ is oscillation peak frequency (determined by average period of the Coulomb oscillation at zero field, as shown in Table S1), $W = 40 \, V^{-1}$ is the integration window, $A(f)$ is the FFT amplitude.}
\label{sfig4}
\end{figure}

\begin{figure}[htb]
\includegraphics[width=1\columnwidth]{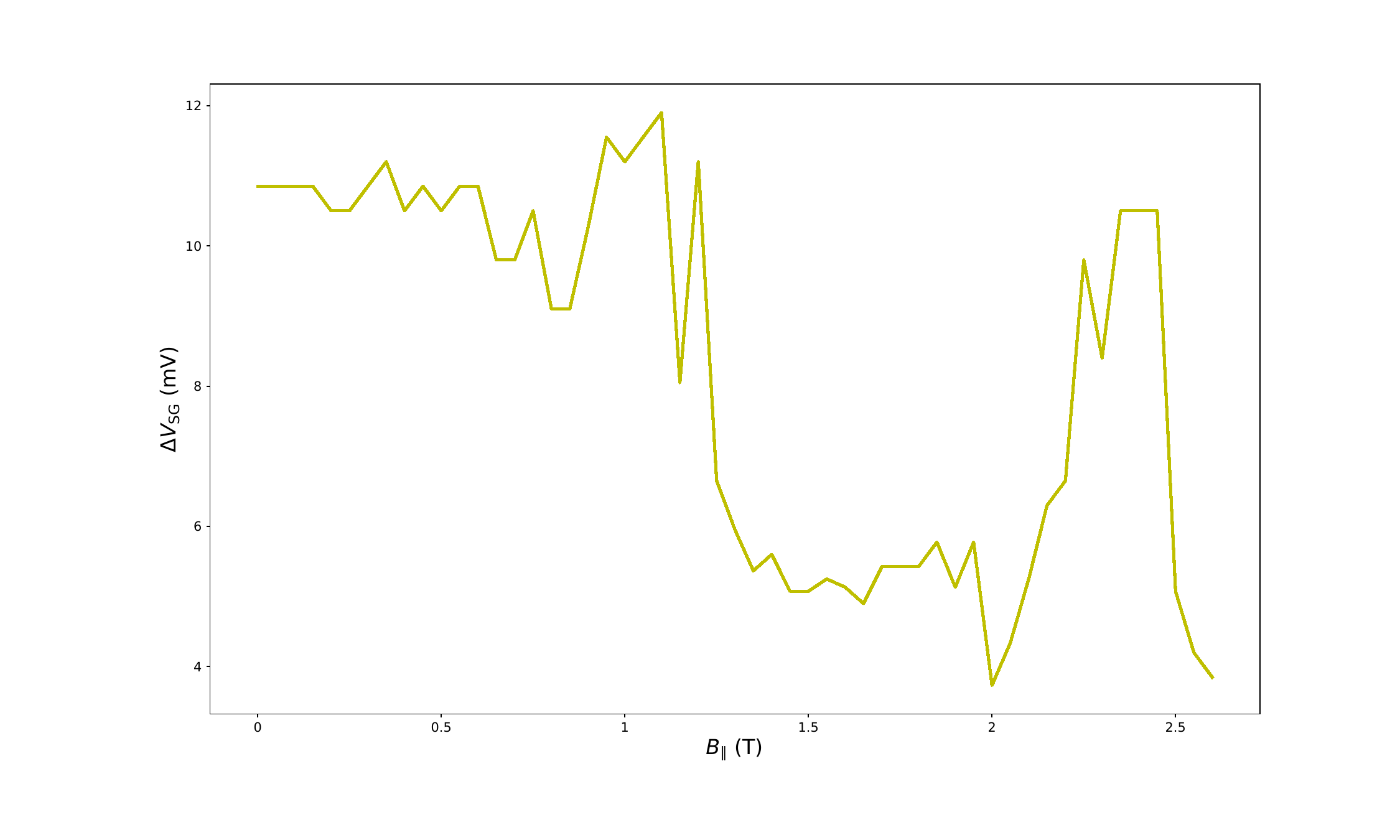}
\centering
\caption{The evolution of the dominated oscillation period as a function of the magnetic field. The data are from Figure 2a, within range $V_\mathrm{SG}$ = [-2.295 V, -2.255 V]. The dominant oscillation period transitions from 2$e$ (10.85 mV) at $B$ = 0 T to 1$e$ (5.1 mV) at $B$ = 1.5 T. To extract the dominated oscillation period, oscillation peak positions within the $V_\mathrm{SG}$ = [-2.295 V, -2.255 V] range under different magnetic fields were extracted using a peak-finding function in Python environment. Peak spacings between adjacent peaks were extracted, then peak spacing data differing from the mean value by more than one standard deviation were excluded. The remaining data's average was recalculated, which is viewed as the dominated oscillation period.}
\label{sfig5}
\end{figure}

\begin{figure}[htb]
\includegraphics[width=1\columnwidth]{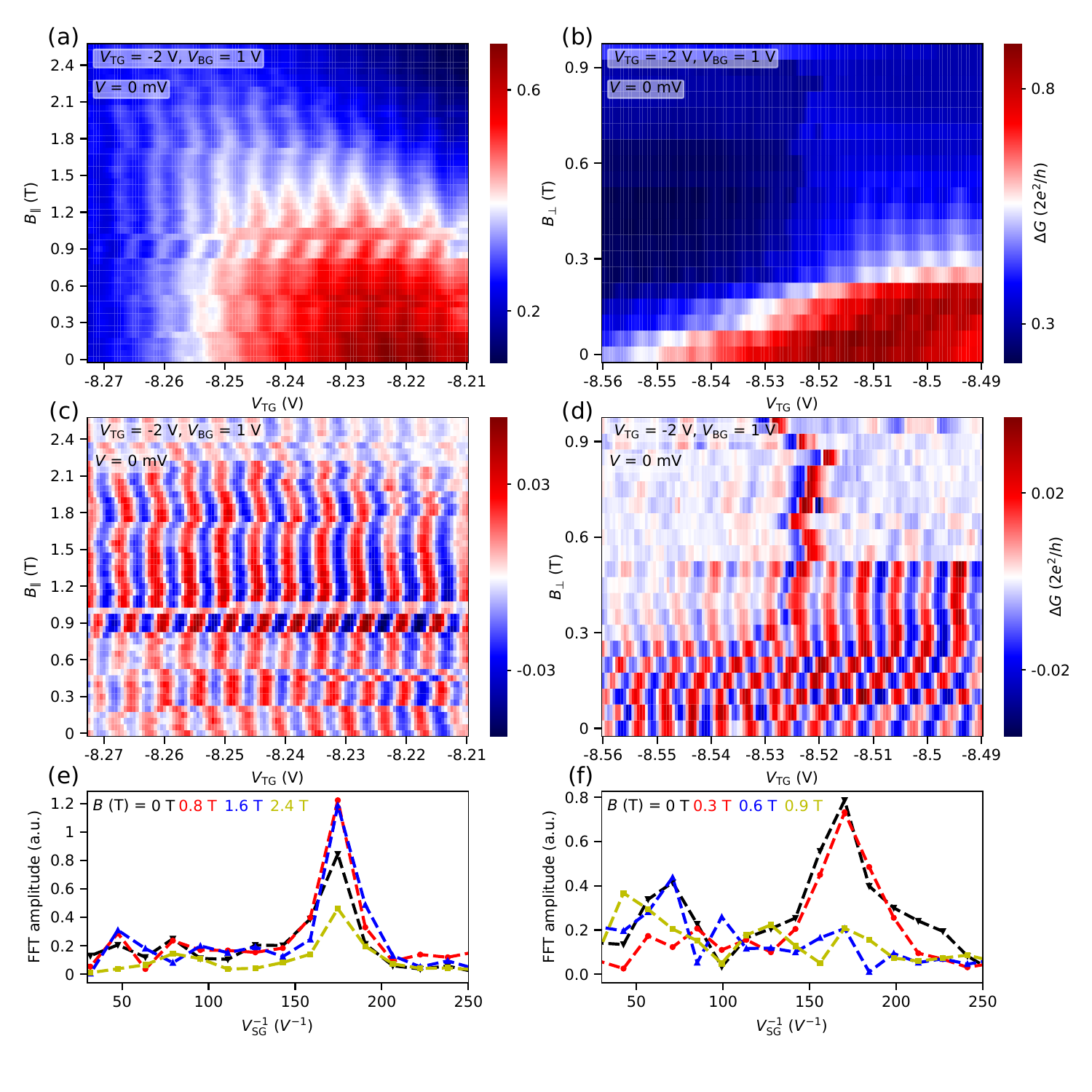}
\centering
\caption{(a.b) 1$e$-period Coulomb blockade oscillation results with parallel and perpendicular magnetic fields of device A. (c,d) Oscillation signal after removing background and (e,f) FFT analysis results.}
\label{sfig6}
\end{figure}

\begin{figure}[htb]
\includegraphics[width=1\columnwidth]{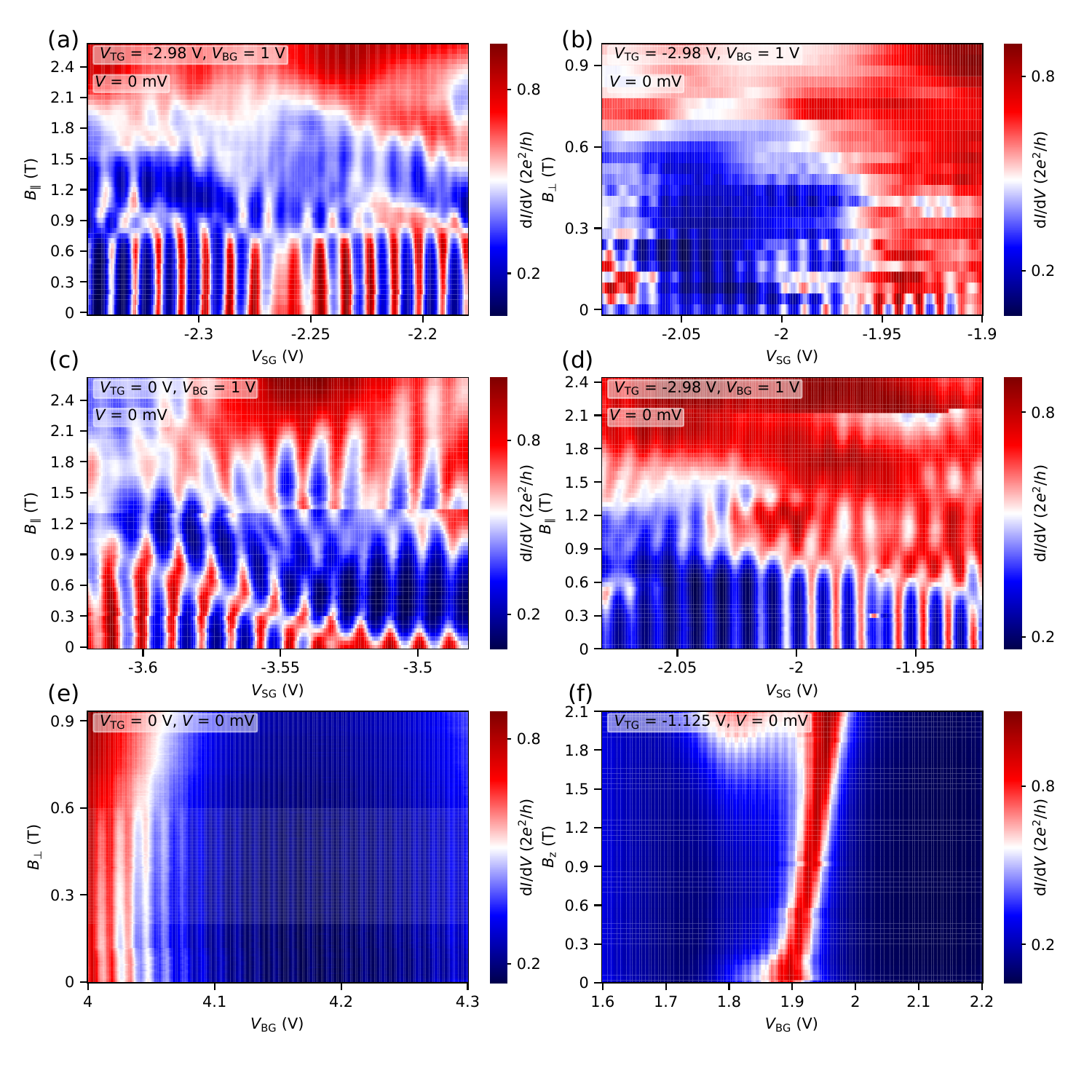}
\centering
\caption{(a,b) Raw data of Fig. 2(a,b) without removing background. (c,d) Additional Coulomb blockade oscillation results with parallel magnetic field of device A. (e) Raw data of Fig. 3(e). (f) Coulomb blockade oscillation results with magnetic field along z-axis of device B. The oscillation period remains unchanged from 0 to 2T.}
\label{sfig7}
\end{figure}
